\documentclass[reprint,aps,amsmath,amssymb,nofootinbib,superscriptaddress, preprintnumbers]{revtex4-2}

\usepackage[utf8]{inputenc}
\usepackage[T1]{fontenc}

\usepackage{amsmath}
\usepackage{mathrsfs}
\usepackage{txfonts}
\usepackage{mathtools}
\usepackage{braket}
\usepackage{tensor}
\usepackage{xcolor}
\usepackage[abbreviations]{siunitx}
\usepackage{graphicx}
\usepackage{booktabs}

\usepackage{float}

\usepackage{braket}
\usepackage{ulem}

\usepackage{hyperref}
\hypersetup{
	colorlinks=true, 
	linktoc=all,     
	linkcolor=blue,  
}

\usepackage{cleveref}

\newcommand{\beq}{\begin{equation}}
\newcommand{\eeq}{\end{equation}}
\newcommand{\beqn}{\begin{eqnarray}}
\newcommand{\eeqn}{\end{eqnarray}}
\newcommand{\bsub}{\begin{subequations}}
\newcommand{\esub}{\end{subequations}}
\newcommand{\bpm}{\begin{pmatrix}}
\newcommand{\epm}{\end{pmatrix}}

\begin{document}
    
 \title{Microscopic study of low-lying states in odd-mass nuclei for atomic electric dipole moment searches}

 \author{E. F. Zhou}   
 \affiliation{School of Physics and Astronomy, Sun Yat-sen University, Zhuhai 519082, P.R. China} 
 
 \affiliation{Guangdong Provincial Key Laboratory of Quantum Metrology and Sensing, Sun Yat-Sen University, Zhuhai 519082, P. R. China }

  \author{J. M. Yao}   
 \email{Contact author: yaojm8@sysu.edu.cn}
  \affiliation{School of Physics and Astronomy, Sun Yat-sen University, Zhuhai 519082, P.R. China}  
  
 \affiliation{Guangdong Provincial Key Laboratory of Quantum Metrology and Sensing, Sun Yat-Sen University, Zhuhai 519082, P. R. China }

\begin{abstract}   
We present a microscopic study of the low-lying states of five odd-mass nuclei of particular interest for experimental searches of atomic electric dipole moments (EDMs): $\nuclide[129]{Xe}$, $\nuclide[199]{Hg}$, $\nuclide[225]{Ra}$, $\nuclide[229]{Th}$, and $\nuclide[229]{Pa}$. The analysis is performed within the recently developed multi-reference covariant density functional theory (MR-CDFT), which incorporates symmetry restoration and configuration mixing based on self-consistent mean-field solutions. The calculated energy spectra and electromagnetic observables of these nuclei are reasonably well reproduced without introducing any parameters beyond those of the underlying universal relativistic energy density functional. The results demonstrate the reliability of MR-CDFT in describing the structure of these nuclei and in providing essential input on nuclear Schiff moments relevant to ongoing EDM searches.

\end{abstract}

\maketitle

\section{Introduction}

The origin of the matter–antimatter asymmetry in the Universe remains one of the most profound open questions in modern physics. As established by Sakharov~\cite{Sakharov:1967}, three conditions must be satisfied to generate a baryon asymmetry in the Universe, including (i) violation of charge conjugation (C) and charge–parity (CP) symmetries, (ii) violation of baryon number conservation, and (iii) departure from thermal equilibrium. Within the Standard Model, CP violation arises from the Cabibbo–Kobayashi–Maskawa (CKM) phase and the QCD $\bar{\theta}$ term, but their contributions are far too small to explain the observed asymmetry. This shortfall has motivated extensive searches for new sources of CP violation beyond the Standard Model, particularly at low-energy scales~\cite{Chupp:2017,Engel:2013review1,Ginges:2003review,Engel:2025}.

Permanent electric dipole moments (EDMs) of nucleons, atoms, and molecules provide exquisitely sensitive probes of such new physics. A nonzero EDM would signal the violation of both parity (P) and time-reversal (T) symmetries and, by the CPT theorem, CP symmetry. Experimental searches have already placed some of the most stringent limits on CP violation, with the current bound on the EDM of $^{199}$Hg atom, $|d_{\text{Hg}}| < 7.4 \times 10^{-30} e \cdot \mathrm{cm}$~\cite{Yoshinaga:2018PTSM3}, representing a sensitivity about three orders of magnitude larger than the Standard Model expectations. The next generation of EDM experiments is expected to achieve even higher precision, potentially opening a new window to physics at the TeV scale and beyond ~\cite{Alarcon:Snowmass2022}.

In diamagnetic atoms such as $^{129}$Xe, $^{199}$Hg, and $^{225}$Ra, atomic EDMs are induced primarily by nuclear Schiff moments, which arise mainly from CP-violating nuclear forces $V_{PT}$ described by P- and T-odd pion–nucleon couplings at leading order in chiral effective field theory~\cite{deVries:2020_FP}. The presence of a nonzero $V_{PT}$ violates the parity of nuclear wave functions and induces a nonzero Schiff moment in the ground state~\cite{Engel:2013PPNP}, which can be evaluated in second-order perturbation theory using the wave functions and energies of nuclear states calculated without $V_{PT}$. Therefore, the calculation of the nuclear Schiff moment requires knowledge of the structural properties of both the nuclear ground state and the excited states with the same spin but opposite parity to the ground state. In particular, the nuclear Schiff moment is proportional to the inverse of the excitation energy of the intermediate state. Thus, in octupole-deformed nuclei such as $^{225}$Ra, the existence of nearly degenerate parity-doublet states enhances the Schiff moment~\cite{Haxton:1983}, whereas in near-spherical nuclei such as $^{129}$Xe and $^{199}$Hg, many higher-lying excitations contribute significantly~\cite{Ban:2010PRC,Yanase:2020LSSM1,Zhou:2025_letter}. Moreover, our study based on beyond-mean-field multireference covariant density functional theory (MR-CDFT) has demonstrated that shape-mixing effects also significantly modify the Schiff moments~\cite{Zhou:2025_letter}. In addition, a strong correlation has been identified between the Schiff-moment contributions and the isovector electric dipole ($E1$) transition strengths, indicating that experimental $B(E1)$ data could help constrain theoretical uncertainties. Therefore, precise knowledge of nuclear structure properties is a prerequisite for achieving an accurate description of nuclear Schiff moments.

In this paper, we present the first MR-CDFT  description of the structural properties of nuclei relevant to the search for atomic EDMs, including $^{129}$Xe, $^{199}$Hg, $^{225}$Ra, $^{229}$Th, and $^{229}$Pa. It is worth noting that all these nuclei are odd-mass nuclei and the collective quadrupole–octupole correlations, which can significantly affect the Schiff moments, need to be carefully taken into account. For odd-mass nuclei with strong octupole deformations, one expects to have parity doublet consisting of a pair of almost degenerate levels of the same spin and opposite parities and connected by a large $E3$ transition matrix element. A microscopic study of such nuclei presents a major challenge for nuclear models. Apart from several phenomenological collective models~\cite{Meyer-Ter-Vehn:1975zck,ADIrving_1979,Helppi:1981doq,Flambaum:1986NPA,Huang:2016srl} and interacting boson–fermion model~\cite{Abu-Musleh:2013uja}, the structural properties of these nuclei have so far been investigated only by the Skyrme Hartree-Fock (SHF) approaches~\cite{Dobaczewski:2005PRL}, the random-phase approximation (RPA)~\cite{deJesus:2005SQRPA1,Ban:2010PRC}, and valence-space shell models~\cite{Yoshinaga:2013PTSM2,Teruya:2017PTSM1,Yanase:2020LSSM1} based on different levels of approximations. With the MR-CDFT framework developed in Ref.~\cite{Zhou:2024PRC}, such a study becomes feasible, starting from a universal energy density functional (EDF) at the beyond-mean-field level and employing the full single-particle basis without introducing any adjustable parameters beyond those of the EDF. This is particularly important for predicting nuclear Schiff moments, which cannot be measured or directly validated experimentally. As demonstrated in this work, we calculate the energy spectra, electromagnetic transition strengths, and electromagnetic moments of low-lying nuclear states and find overall reasonable agreement with the available experimental data, thereby providing theoretical support for the Schiff-moment calculations reported in Ref.~\cite{Zhou:2025_letter}.
 
The paper is organized as follows. Section~\ref{sec:framework} briefly outlines the theoretical framework employed to describe nuclear low-lying states. The results and their discussion are presented in Sec.~\ref{sec:results}, and the main conclusions are summarized in Sec.~\ref{sec:summary}.

\section{Theoretical framework}
\label{sec:framework}

The MR-CDFT for the low-lying states of odd-mass nuclei has been introduced in detail in Ref.~\cite{Zhou:2024PRC}. Here, we only present a brief description of this theory, in which the wave functions of low-lying states are constructed as a mixing of configurations with different deformation parameter $\mathbf{q}$ and quantum number $K$,  
\begin{equation}
\label{eq:gcmwf}
\vert \Psi^{J\pi}_\alpha\rangle
=\sum_{c} f^{J\alpha \pi}_{c}  \ket{NZ J\pi; c}.
\end{equation} 
Here, $c$ is a collective label for $(K,\mathbf{q})$, and $\alpha$ distinguishes states with the same angular momentum $J$. The basis function with quantum numbers ($NZJ\pi$) is given by
\begin{equation}
\label{eq:basis}
\ket{NZ J\pi; c} 
=  \hat P^J_{MK} \hat{P}^\pi \hat P^N\hat P^Z \ket{\Phi^{\rm (OA)}_\kappa(\mathbf{q})},
\end{equation}
where $\hat P^{J}_{MK}$,$\hat{P}^\pi$ and $\hat{P}^{N(Z)}$  are projection operators that select components with the angular momentum $J$, parity $\pi$, and neutron(proton) number $N(Z)$~\cite{Ring:1980}. 

The mean-field configurations $\ket{\Phi^{\mathrm{(OA)}}_\kappa(\mathbf{q})}$ for odd-mass nuclei are constructed as one-quasiparticle (1qp) states,
\begin{eqnarray}
\label{eq:odd-mass-wfs}
\ket{\Phi^{\mathrm{(OA)}}_\kappa(\mathbf{q})} = \alpha^\dagger_\kappa \ket{\Phi_{(\kappa)}(\mathbf{q})},
\end{eqnarray}
where $\ket{\Phi_{(\kappa)}(\mathbf{q})}$ denotes a quasiparticle vacuum state with even number parity (but constrained to have an odd average particle number) obtained using the false quasiparticle vacuum (FQV) scheme~\cite{Bally:2014,Zhou:2024PRC} within the single-reference (SR) CDFT framework.  This framework starts from the  relativistic EDF composed of the standard kinetic energy, nucleon-nucleon ($NN$) interaction energy, as well as the electromagnetic energy terms~\cite{Burvenich:2002PRC,Zhao:2010PRC},
 \beqn
 \label{eq:EDF}
  &&  E_{\rm EDF}[\tau, \rho, \nabla\rho; \mathbf{C}] \nonumber\\
 &=& \int d^3 r \Bigg[ 
   \sum_k \varv_k^2\psi_k^\dag\left(\boldsymbol{\alpha}\cdot\boldsymbol{p}+\beta m_N-m_N\right)\psi_k \nonumber\\
   &&+\cfrac{\alpha_S}{2}~\rho_S^2 +\cfrac{\beta_s}{3}~\rho_S^3
     +\cfrac{\gamma_S}{4}~\rho_S^4 +\cfrac{\delta_S}{2}~\rho_S\nabla^2 \rho_S \nonumber\\
     &&+\cfrac{\alpha_V}{2}~(j_V)_\mu j^\mu_V + \cfrac{\gamma_V}{4}~[(j_V)_\mu j^\mu_V]^2 +\cfrac{\delta_V}{2}~(j_V)_\mu \nabla^2j^\mu_V\nonumber\\
     && + \cfrac{\alpha_{TV}}{2}~\vec{j}^\mu_{TV} (\vec{j}_{TV})_\mu
     + \cfrac{\delta_{TV}}{2} \vec{j}^\mu_{TV} \nabla^2(\vec{j}_{TV})_{\mu}\nonumber\\
     &&+ \cfrac{1}{4}~F_{\mu\nu}F^{\mu\nu}-F_{0\mu}\partial_0 A_{\mu} + e A_\mu j_\rho^{~\mu} 
   \Bigg]. 
\eeqn
 The low-energy parameters in the $NN$ interaction energy are collectively denoted as $\mathbf{C}=\{\alpha_S, \beta_S, \gamma_S, \delta_S, \alpha_V, \gamma_V, \delta_V, \alpha_{TV}, \delta_{TV}\}$. The subscripts $(S, V)$ indicate the scalar and vector types of $NN$ interaction vertices in Minkowski space, respectively, and $T$ for the vector in isospin space.  The PC-PK1 parameterization~\cite{Zhao:2010PRC} is employed in this work. These parameters were optimized to reproduce the ground-state bulk properties of dozens of spherical nuclei and the properties of infinite nuclear matter around saturation density. The same set of parameters is used for all nuclei across the nuclear chart and has proven to be highly successful in describing nuclear masses and other observables~\cite{DRHBcMassTable:2024}.
 
 The densities and currents in the relativistic EDF (\ref{eq:EDF}), denoted collectively as $\rho$, are defined by    
\beqn
  \rho_S(\boldsymbol{r}) &=& \sum_k\varv_k^2\bar \psi_k(\boldsymbol{r})\psi_k(\boldsymbol{r}),\\
     j_V^{~\mu}(\boldsymbol{r}) &=& \sum_k\varv_k^2 \bar \psi_k(\boldsymbol{r}) \gamma^\mu\psi_k(\boldsymbol{r}),\\
     \vec{j}_{TV}^\mu(\boldsymbol{r}) &=& \sum_k\varv_k^2 \bar \psi_k(\boldsymbol{r}) \vec{\tau} \gamma^\mu\psi_k(\boldsymbol{r}).
\eeqn
In the above summations, single-particle states in the Dirac sea are excluded. The quantity $\varv_k^2$ denotes the occupation probability of the $k$th single-particle state, whose wave function is represented by the Dirac spinor $\psi_k(\boldsymbol{r})$.

The quasiparticle creation operator $\alpha^\dagger_\kappa$ in (\ref{eq:odd-mass-wfs}) changes the number parity of a mean-field state $\ket{\Phi_{(\kappa)}(\mathbf{q})}$ from even to odd, and the index $\kappa$ distinguishes different quasiparticle configurations. In the present study, axial symmetry is assumed, and $\mathbf{q}=\{\beta_{2}, \beta_3\}$ labels the quadrupole and octupole deformation parameters,
\begin{equation}
    \beta_\lambda = \frac{4\pi}{3AR^\lambda} \bra{\Phi_{(\kappa)}(\mathbf{q})}\hat Q_{\lambda 0}\ket{\Phi_{(\kappa)}(\mathbf{q})},\quad R=1.2 A^{1/3}~ {\rm fm},
\end{equation}
with $A$ denoting the nuclear mass number. Each configuration is therefore labeled by the quantum numbers $K$ (the projection of nuclear total angular momentum along the body-fixed 3 axis), which should equal to the 3-component $\Omega$ of the angular momentum for the unpaired nucleon, i.e., $K = \Omega$~\cite{Ring:1980}. Unless otherwise specified, only the lowest-energy quasiparticle state is considered in the FQV scheme~\cite{Zhou:2024PRC}.

The weight function $f^{J\alpha \pi}_{c}$ in Eq.~(\ref{eq:gcmwf}) is determined by the variational principles, leading to the following Hill-Wheeler-Griffin (HWG) equation~\cite{Hill:1953,Ring:1980},
\begin{eqnarray}
\label{eq:HWG}
\sum_{c'}
\Bigg[\mathscr{H}^{NZJ\pi}_{cc'}
-E_\alpha^{J\pi }\mathscr{N}^{NZJ\pi }_{cc'} \Bigg]
f^{J\alpha \pi}_{c'}=0,
\end{eqnarray}
where the Hamiltonian kernel  and norm kernel are defined by
\begin{equation}
\label{eq:kernel}
 \mathscr{O}^{NZJ\pi}_{cc'}
 =\bra{NZ J\pi; c}  \hat O \ket{NZ J\pi; c'},
\end{equation}
with the operator $\hat O$ representing $\hat H$ and $1$, respectively. In the present study based on the covariant EDF (\ref{eq:EDF}),  the mixed-density prescription is employed in the evaluation of the Hamiltonian kernel~\cite{Bonche:1990NPA,Robledo:2018JPG}. Details on the calculation of the kernels in Eq.~(\ref{eq:kernel}) can be found in Ref.~\cite{Zhou:2024PRC}.

The HWG equation (\ref{eq:HWG}) for a given set of quantum numbers $(NZJ\pi)$ is solved in the standard way, as discussed in Refs.~\cite{Ring:1980,Yao:2010}. This is done by first diagonalizing the norm kernel $\mathscr{N}^{NZJ\pi }_{cc'}$. A new set of basis functions is then constructed using the eigenfunctions of the norm kernel with eigenvalues larger than a pre-chosen cutoff value, which removes possible redundancy in the original basis.  The Hamiltonian is diagonalized on this new basis. In this way, the energies $E_\alpha^{J\pi}$ and
the mixing weights $f^{J\alpha\pi}_{c}$ of the nuclear states $\vert \Psi^{J\pi}_\alpha\rangle$ can be obtained. Since the basis functions $\ket{NZ J\pi; c}$ are nonorthogonal to each other, one usually introduces the collective wave function $g^{J\pi}_\alpha(K,\mathbf{q})$ as shown below
\begin{equation}
\label{eq:coll_wf}
g^{J\pi}_\alpha(K,\mathbf{q})=\sum_{c'} (\mathscr{N}^{1/2})^{NZJ\pi}_{c,c'} f^{J\alpha\pi}_{c'},
 \end{equation}
 which fulfills the normalization condition. The distribution of $g^{J\pi}_\alpha(K,\mathbf{q})$ over $K$ and $\mathbf{q}$ reflects the contribution of each basis function to the nuclear state $\vert \Psi^{J\pi}_\alpha\rangle$.
 
 With the mixing weight $f^{J\alpha\pi}_{c}$, it is straightforward to determine the observables of nuclear low-lying states, including the electric quadrupole moment $Q_s$, the magnetic dipole moment $\mu$, as well as the $E2$ and $M1$ transition strengths. 
 The strength of the $E\lambda (M\lambda)$ transition from the initial state $\ket{\Psi^{J_i\pi_i}_{\alpha_i}}$ to the final state $\ket{\Psi^{J_f\pi_f}_{\alpha_f}}$ is determined by 
\begin{eqnarray}
&& B(T\lambda,J_i\alpha_i\pi_i\rightarrow J_f\alpha_f\pi_f) = \cfrac{1}{2J_i+1} \nonumber\\
&&\times\left| \sum_{c_f, c_i} f^{J_i\alpha_i \pi_i}_{c_i}f^{J_f\alpha_f \pi_f}_{c_f} 
    \langle NZJ_f\pi_f,c_f||\hat T_\lambda ||NZJ_i\pi_i,c_i\rangle
    \right|^2,
\end{eqnarray} 
where the configuration-dependent reduced matrix element is simplified as follows, 
\begin{eqnarray}
\label{eq:reduced_matrix_element}
    &&\langle NZ J_f\pi_f; c_f ||\hat  T_\lambda|| NZ J_i\pi_i; c_i\rangle\nonumber\\
    =&& \delta_{\pi_f\pi_i,(-1)^\lambda}(-1)^{J_f-K_f}
    \cfrac{\hat{J}_i^2\hat{J}_f^2}{8\pi^2} \sum_{\nu M} \left(
 \begin{array}{ccc}
J_f&\lambda&J_i\\
-K_f &\nu&M\\
\end{array}
    \right) \nonumber\\
   &&\times  \int d\Omega D_{MK_i}^{J_i\ast}(\Omega)
\bra{\Phi^{\rm (OA)}_{\kappa_f}(\mathbf{q}_f)}
    \hat T_{\lambda\nu}
  \hat R(\Omega) \hat P^Z\hat P^N\hat P^{\pi_i}
   \ket{\Phi^{\rm (OA)}_{\kappa_i}(\mathbf{q}_i)},\nonumber \\
\end{eqnarray}
where $\hat T_{\lambda\nu}$ represents an $E\lambda(M\lambda)$ operator and $\hat J=\sqrt{2J+1}$. 
The magnetic dipole moment $\mu(J_\alpha^\pi)$ of the state  $\ket{\Psi^{J\pi}_{\alpha}}$   can be calculated by 
\begin{eqnarray}
\label{eq:magnetic_mu}
    \mu(J_\alpha^\pi) 
    &=& \bra{\Psi^{J\pi}_\alpha}\hat \mu_{10}\ket{\Psi^{J\pi}_\alpha}\nonumber\\
   &=& \left(
     \begin{array}{ccc}
J&1&J\\
-J&0&J\\
\end{array}
   \right)
  \sum_{c_f, c_i} f^{J_i\alpha_i \pi_i}_{c_i} f^{J_f\alpha_f \pi_f}_{c_f} \nonumber\\
   &&\times 
   \langle NZ J_f\pi_f; c_f ||\hat  \mu_1|| NZ J_i\pi_i; c_i\rangle.
\end{eqnarray}
The magnetic dipole operator $\hat  \mu_{1\mu}$ of rank-one tensor can be constructed by the vector operator $\hat{\boldsymbol{\mu}}$, which is determined by the current operator,
 \begin{eqnarray}
        \hat{\boldsymbol{\mu}}
        &=& \int d^3 \boldsymbol{r}\left[\frac{m_Nc^2}{\hbar c}e\psi^{\dagger}(\boldsymbol{r}) \mathbf{r} \times \boldsymbol{\alpha} \psi(\boldsymbol{r})
        +\kappa \psi^{\dagger}(\boldsymbol{r}) \beta \boldsymbol{\Sigma} \psi(\boldsymbol{r})\right],\nonumber\\
 \end{eqnarray}
with $e$ being the electric charges of nucleons,  $\kappa$ the free-space anomalous gyromagnetic ratio (called $g$ factor) of the nucleon, i.e., $\kappa_p=1.793$ for protons, $\kappa_n=-1.913$ for neutrons, and $m_N$ the mass of nucleon. The Dirac gamma matrices read
\beqn 
\boldsymbol{\alpha}=\begin{pmatrix}
0 & \boldsymbol{\sigma} \\
 \boldsymbol{\sigma} & 0
\end{pmatrix},\quad \beta \boldsymbol{\Sigma} =\begin{pmatrix}
\boldsymbol{\sigma} & 0\\
0 & -\boldsymbol{\sigma}
\end{pmatrix}.
\eeqn 
Here, $\boldsymbol{\sigma}$ is the Pauli spinor. The units of the magnetic dipole moment $\hat{\boldsymbol{\mu}}$ is nuclear magneton ($\mu_N=e\hbar/m_N$).

 The spectroscopic quadrupole moment $Q^{\rm s}(J^\pi_\alpha)$ of the state $\ket{\Psi^{J\pi}_{\alpha}}$ is given by
\begin{eqnarray}
\label{eq:spectroscopic_Q}
    Q^{\rm s}(J^\pi_\alpha) 
    &=&  \sqrt{\cfrac{16\pi}{5}}\bra{\Psi^{J\pi}_\alpha}\hat Q_{20}\ket{\Psi^{J\pi}_\alpha}\nonumber\\
    &=&  \sqrt{\cfrac{16\pi}{5}}
    \left(
     \begin{array}{ccc}
J&2&J\\
-J&0&J\\
\end{array}
   \right)
  \sum_{c_f, c_i} f^{J_i\alpha_i \pi_i}_{c_i} f^{J_f\alpha_f \pi_f}_{c_f} \nonumber\\
   &&\times 
   \langle NZ J_f\pi_f; c_f ||er^2Y_{2}|| NZ J_i\pi_i; c_i\rangle.
\end{eqnarray}
The reduced transition matrix elements in (\ref{eq:magnetic_mu}) and (\ref{eq:spectroscopic_Q}) are determined  by Eq.(\ref{eq:reduced_matrix_element}). 
It is worth noting that, since the transition operators are defined in the full single-particle space in the MR-CDFT calculations, the bare electric charges of protons and neutrons, $e_p = 1$ and $e_n = 0$, are employed.

\section{Results and discussion}
\label{sec:results}

In this work, the mean-field wave functions of the odd-mass nuclei are calculated self-consistently within the SR-CDFT framework, where the blocked single-particle orbitals are treated using the FQV scheme~\cite{Zhou:2024PRC} while preserving time-reversal symmetry. The Dirac equation for the single-particle wave functions, $\psi_k$, is solved in a three-dimensional isotropic harmonic-oscillator basis including 12 major shells, which is sufficient to achieve reasonable convergence for spectroscopic quantities. Pairing correlations between nucleons are taken into account using the BCS method with a zero-range pairing force~\cite{Zhao:2010hi}. The numbers of mesh points for the rotational Euler angle $\beta$ in the angular-momentum projection (AMP) and for the gauge angle $\varphi$ in the particle-number projection (PNP) are set to $N_\beta = 16$ and $N_\varphi = 7$, respectively, in the calculation of the projected Hamiltonian and norm kernels. In the configuration-mixing GCM calculations, the convergence of the results with respect to the number of natural states and the selection of configurations has been carefully examined. In this work,  only the states with angular momentum $J\leq 15/2$ are calculated with the MR-CDFT. Further details of the MR-CDFT calculations for odd-mass nuclei can be found in Ref.~\cite{Zhou:2024PRC}.

\subsection{\nuclide[129]{Xe}}

The low-lying states of $^{129}$Xe have been investigated with several different types of nuclear models, including the particle-rotor model (PRM)~\cite{Meyer-Ter-Vehn:1975zck},  the particle–vibration coupling model~\cite{ADIrving_1979,Huang:2016srl}, core–quasiparticle coupling model~\cite{Helppi:1981doq},  interacting boson–fermion model~\cite{Abu-Musleh:2013uja} and cranking shell-model calculations~\cite{Huang:2016srl}.   In the following, we present the results of MR-CDFT study of the low-lying states in $^{129}$Xe, which explicitly incorporates shape fluctuations in both quadrupole and octupole shapes degrees of freedom.

\begin{figure}[]
 \centering
\includegraphics[width=8.5cm]{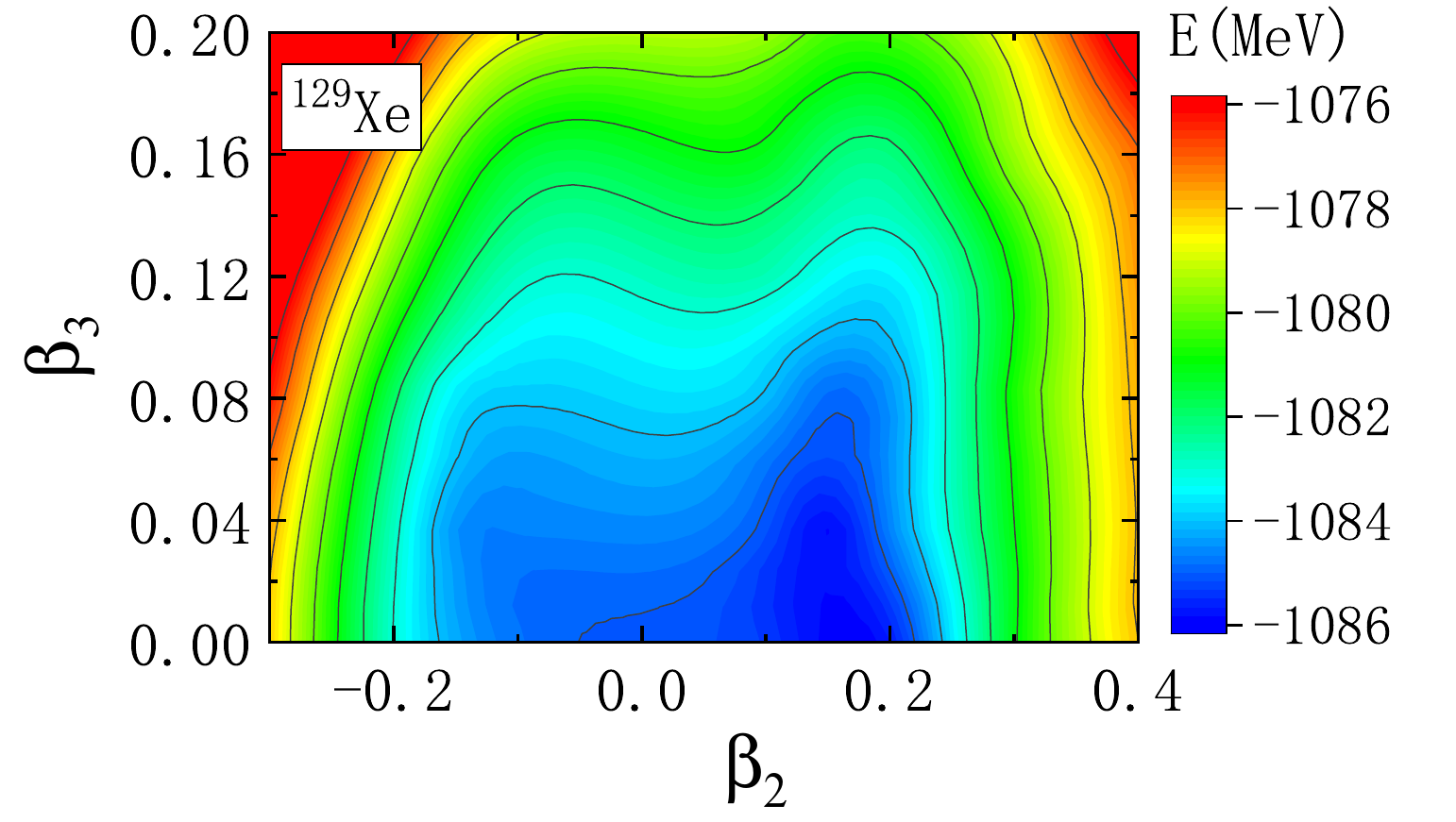}
\caption{Mean-field energy surface of \nuclide[129]{Xe} calculated with the FQV scheme in the quadrupole–octupole ($\beta_2, \beta_3$) deformation plane.}
 \label{fig:Xe129_MFE}
 \end{figure}
 
 \begin{figure}[]
 \centering
\includegraphics[width=8.5cm]{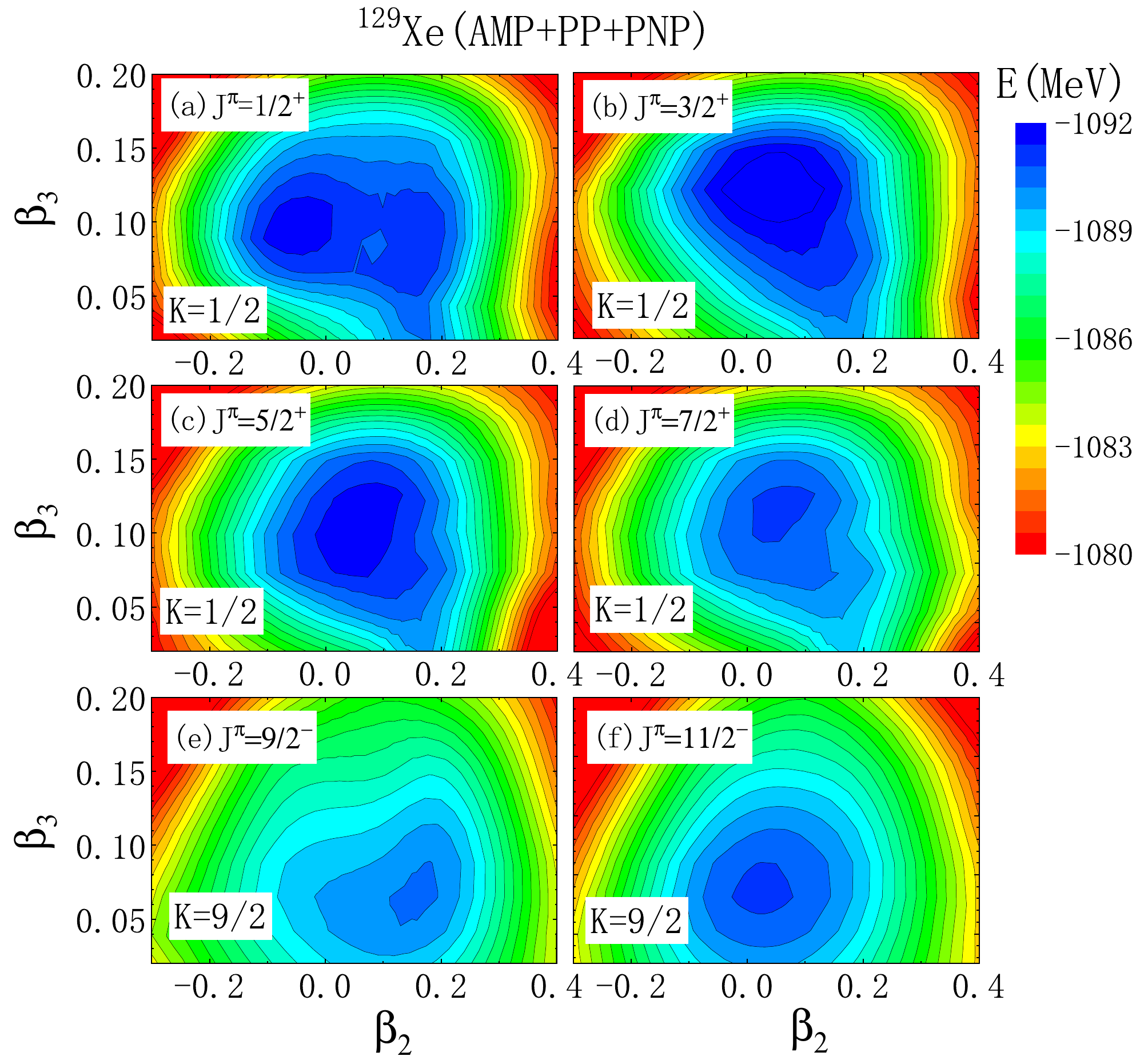}
\caption{ Energy surfaces of quantum-number projected states in the $(\beta_2, \beta_3)$ deformation plane for $^{129}$Xe, with simultaneous projection onto spin–parity $(J^\pi)$, and particle numbers $(N, Z)$. The $K$ quantum numbers of the configurations are indicated for each state.
}
 \label{fig:Xe129_PE}
 \end{figure}

Figure~\ref{fig:Xe129_MFE} displays the mean-field potential energy surface (PES) of $^{129}$Xe in the $(\beta_2, \beta_3)$ deformation plane. One observes that the  PES exhibits a weakly prolate energy minimum  with $\beta_2 \approx 0.16$, $\beta_3=0$, even though it is soft along both $\beta_2$ and $\beta_3$ directions. The $\beta_2$ value of the energy minimum is close to the value 0.18 found in the previous study using a Skyrme-type pseudo-potential~\cite{Bally:2022rhf}.
 The energies of states for $^{129}$Xe with projections onto nucleon numbers $(N, Z)$ and different spin parities $J^\pi$ are shown in Fig.~\ref{fig:Xe129_PE}. The quasiparticle configurations are labeled by the quantum number $K$. For each $K$, we only show the energy of the lowest-energy quasiparticle configuration in the $(\beta_2,\beta_3)$ deformation plane.  It is seen that the projected state with the lowest energy corresponds to the $J^\pi=1/2^+$ state, built on the configuration with $K=1/2$. Its energy is nearly degenerate with that of the $J^\pi=3/2^+$ state projected from the same configuration, with the energies of the global minimum being $-1091.58$ MeV and $-1091.11$ MeV, respectively. The $J^\pi=1/2^+$ state is associated with a weakly oblate deformation ($\beta_2 \approx -0.08$), while the $J^\pi=3/2^+$ state corresponds to a weakly prolate deformation ($\beta_2 \approx 0.10$); both exhibit an octupole-deformed minimum at $\beta_3 \approx 0.1$. Compared with the mean-field results, the octupole correlations become more pronounced after projections, whereas the quadrupole deformation appears to be somewhat reduced. A more detailed comparison in Figs.~\ref{fig:Xe129_PE}(a–d) shows that, as the angular momentum increases, the energy minimum gradually shifts toward prolate shapes and the nucleus becomes increasingly rigid.

Figures~\ref{fig:Xe129_PE}(e) and (f) present the quantum-number–projected PESs based on the configurations with $K= 9/2$. The energy minimum of the $J^\pi = 11/2^-$ state is found at a smaller quadrupole deformation ($\beta_2 = 0.03$), with an energy lower than that of the $J^\pi = 9/2^-$ state. This behavior can be understood within the PRM in the weak-coupling limit~\cite{Ring:1980}, where the lowest state corresponds to  $J = j$, with $j$ being the single-particle angular momentum. In the present case, the blocked orbital originates from the spherical $h_{11/2}$ subshell with $\Omega = 9/2$. Moreover, the octupole deformation at the energy minimum of the $J^\pi = 1/2^+$, $3/2^+$, $5/2^+$, and $7/2^+$ states is found to be around or even larger than $\beta_3 = 0.1$, much greater than that of the projected $J^\pi = 9/2^-$ and $11/2^-$ states, again indicating the presence of rather strong octupole correlations in the positive-parity states.

 \begin{figure*}[]
 \centering
\includegraphics[width=0.7\paperwidth]{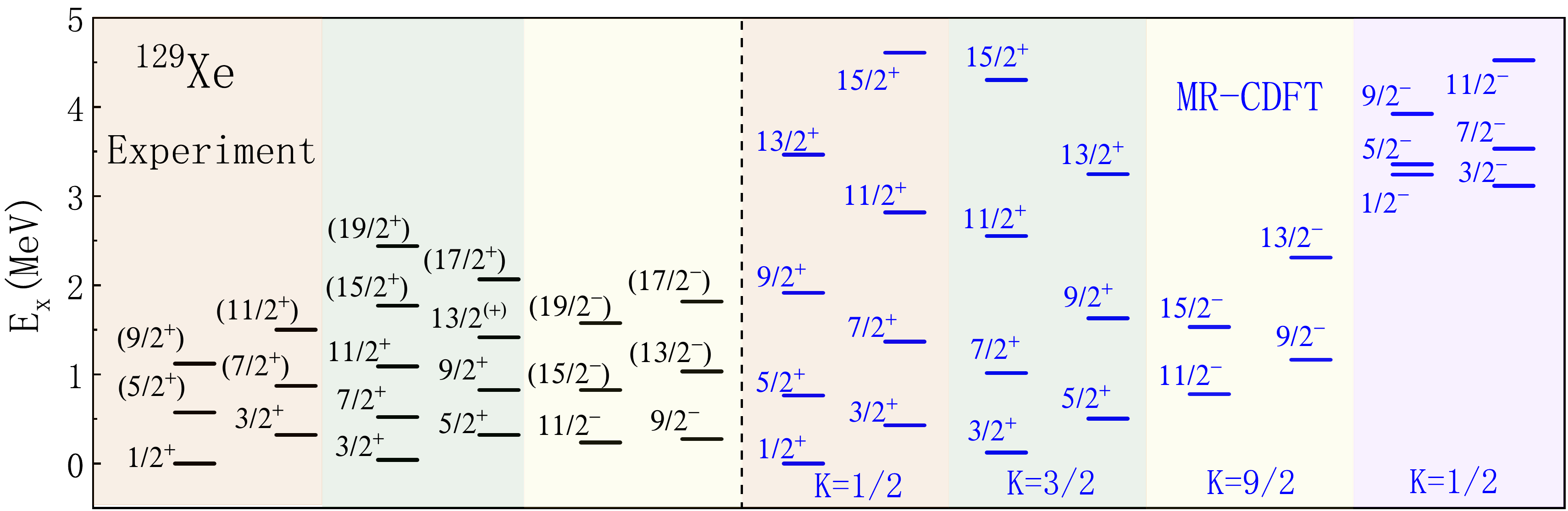}
\caption{Comparison of the energy spectra of $^{129}$Xe calculated with MR-CDFT and the available data from Ref.~\cite{NNDC}.}
 \label{fig:Xe129_spectra}
 \end{figure*}

    \begin{figure}[]
 \centering
\includegraphics[width=8.5cm]{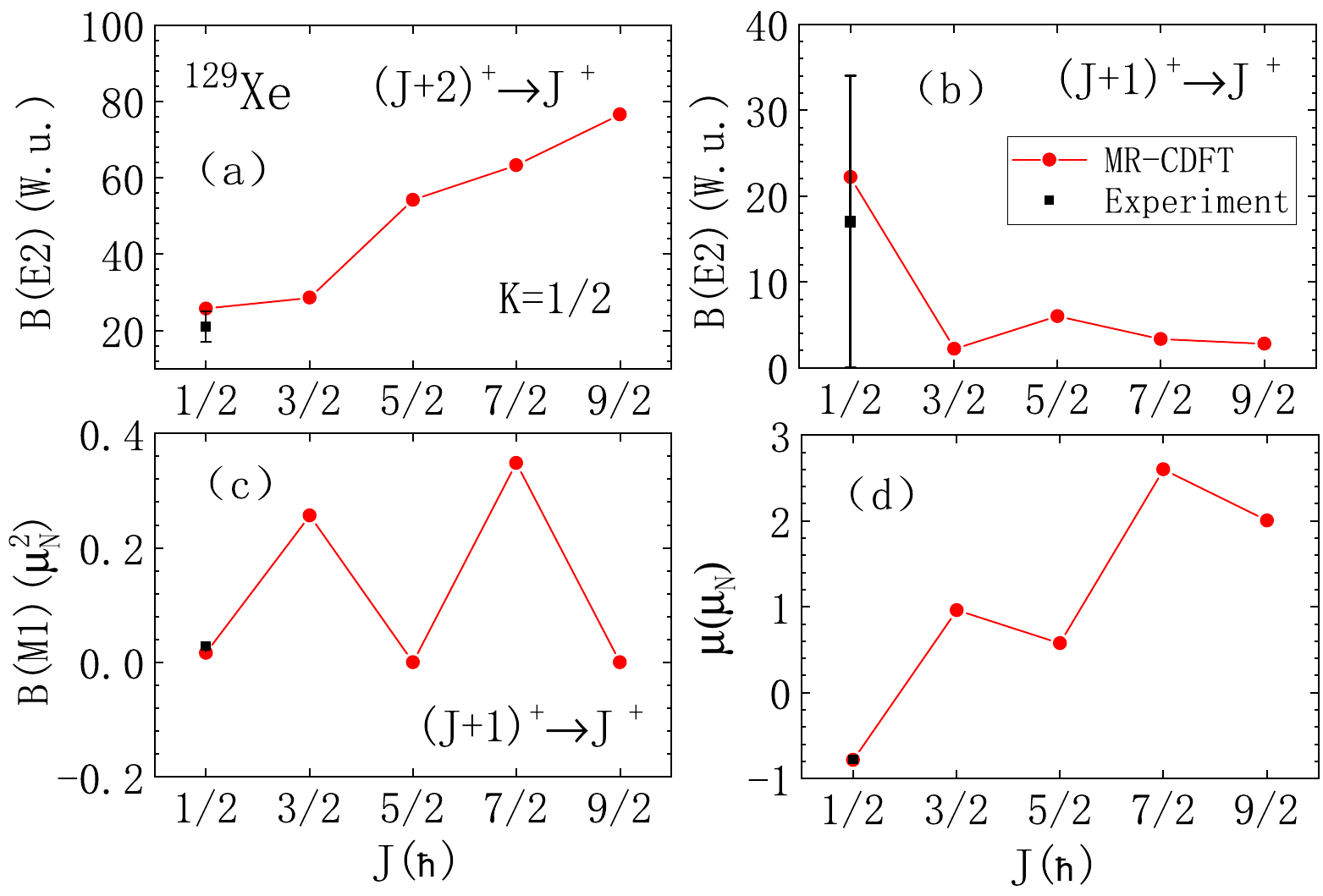}
\caption{Electromagnetic properties of the low-lying states of $^{129}$Xe with $K=1/2$ obtained from the MR-CDFT calculation, in comparison with available data. (a) Electric quadrupole ($E2$) transition strengths for $J+2 \rightarrow J$; (b) $E2$ transition strengths for $J+1 \rightarrow J$; (c) magnetic dipole ($M1$) transition strengths for $J+1 \rightarrow J$; and (d) magnetic dipole moments of states as functions of $J$. Experimental data from Ref.~\cite{NNDC} are shown for comparison.}
 \label{fig:Xe129_trans}
 \end{figure}

 Figure \ref{fig:Xe129_spectra} compares the calculated low-lying spectrum of $^{129}$Xe from MR-CDFT with experimental data. Overall, the main features of the spectrum are reproduced well. Specifically, the spin parity $1/2^+$ of the ground state is reproduced. According to our study, the ground state is dominated by an admixture of the neutron configurations $\Omega[Nn_zm_\ell] = \nu1/2[400]$ and $\nu1/2[440]$, where $N$ denotes the principal quantum number and $m_\ell$ is the projection of the orbital angular momentum onto the symmetry axis, with $\Omega = m_\ell + m_s$~\cite{Ring:1980}. The predicted ground-state magnetic moment $\mu(1/2^+)=-0.785~\mu_N$  is close to the data $-0.778~\mu_N$~\cite{NNDC}.  If neglecting the mixing of configurations with octupole shapes, the magnetic moment becomes $-1.65~\mu_N$, significantly deviating from the data. With the octupole shape fluctuation, we also obtain a series of negative-parity states dominated by the configuration $\nu1/2[550]$, with the bandhead $1/2^-$ state  located around 3.24 MeV. Although these states have not yet been observed experimentally, their predicted positions, particularly that of the $1/2^-$ state,  play a crucial role in determining the nuclear Schiff moment.
 
 Moreover, the coexistence of two sets of rotational bands built on the $3/2^+$ and $11/2^-$ states, respectively, is successfully described, although the calculated levels are systematically stretched in energy. In particular, our result shows that the bands on top of the $3/2^+$ and $5/2^+$ states belong to the same band but with  opposite signatures. This rotational band is dominated by the configuration of $\nu 3/2[402]$, consistent with the conclusion in Ref.~\cite{Huang:2016srl}.  
 The electromagnetic properties of the low-lying states of $^{129}$Xe are also well captured, as shown in Fig.~\ref{fig:Xe129_trans}, which displays the $E2$ and $M1$ transition strengths and magnetic dipole moments for the ground-state band with $K = 1/2$. The calculated $B(E2)$ values increase with angular momentum for $\Delta J = 2$ transitions and decrease for $\Delta J = 1$ transitions, with a minor deviation observed in the $5/2^+ \rightarrow 3/2^+$ transition. Generally, the weak $B(E2)$ transitions suggest that $^{129}$Xe is nearly spherical or weakly deformed.  
 
 We note that the $11/2^-$ state could be dominated by either of the quasiparticle configurations  $\nu 11/2[505]$ or $\nu 9/2[514]$, which lie close in energy. The magnetic moment provides a sensitive probe to the configuration of nuclear state.  To determine the predominant configuration, we calculate the magnetic moment of the $11/2^-$ state based on the configuration  $\nu 9/2[514]$ and obtain $\mu(11/2^-) = -0.78~\mu_N$, in reasonable agreement with the experimental  value of $-0.891223(4)~\mu_N$. In contrast, when the configuration $\nu 11/2[505]$ with $K = 11/2$  is employed, the calculated magnetic moment decreases to $\mu = -0.42~\mu_N$.  This indicates that the predominant configuration of the $11/2^-$ state is  $\nu 9/2[514]$ rather than $\nu 11/2[505]$.  This conclusion differs from the assignment proposed in Ref.~\cite{Huang:2016srl}.
 
 The staggering behaviors in the predicted $B(M1)$ and magnetic moments of Fig.~\ref{fig:Xe129_trans}(c) and (d) are a typical phenomenon for bands with $K = 1/2$. It can be understood from the PRM  in which, the $g$ factor of the state $(J^\pi)$ of the rotational band with $K=1/2$  is given by~\cite{Bohr:1998v2,Stuchbery:2000}
 \begin{equation}
g^{\rm PRM} = g_R + \frac{K^2}{J(J+1)} (g_K - g_R)
\left[ 1 +  (2J + 1)(-)^{J + 1/2} b_0 \right],
\end{equation}
where $g_R$ is the $g$-factor of the rotational core, and $g_K$ is the intrinsic $g$-factor associated with the single nucleon.
The $b_0$ is the magnetic decoupling parameter originated from the Coriolis matrix elements~\cite{Bohr:1998v2}, which gives rise to the signature-dependent staggering behavior for bands with $K = 1/2$.  

 \begin{figure}[]
 \centering
\includegraphics[width=8.5cm]{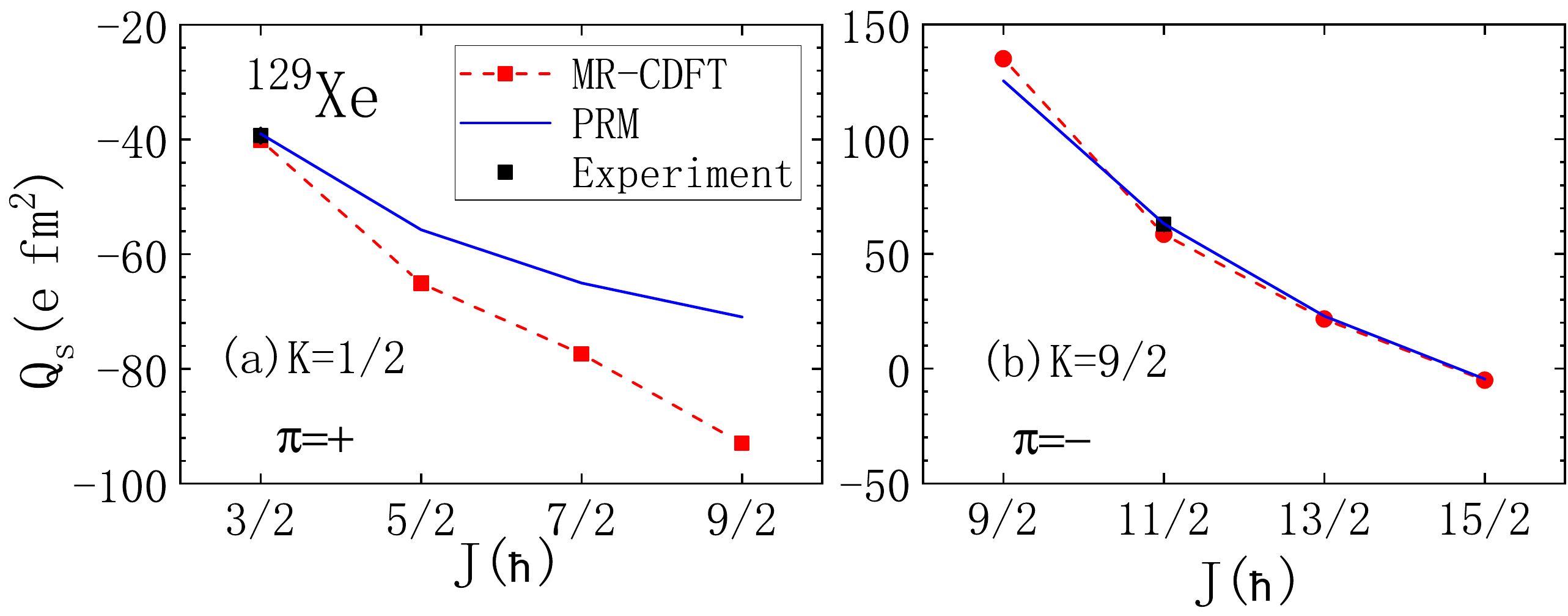}
\caption{
Spectroscopic quadrupole moments of $^{129}$Xe for the positive-parity ($K = 1/2$) (a) and negative-parity ($K = 9/2$) (b) states  calculated with the MR-CDFT, in comparison with the PRM results and available  data~\cite{NNDC}.
 }
 \label{fig:Xe129_qs}
 \end{figure}

Figure~\ref{fig:Xe129_qs} displays the spectroscopic quadrupole moments $Q_s$ of the positive and negative-parity states of $^{129}$Xe with $K = 1/2$ and $K = 9/2$, respectively, from the MR-CDFT calculation, in comparison with the available data. It is seen that the  $Q_s(3/2^+) = -40.1~e\,\mathrm{fm}^2$ and $Q_s(11/2^-) = 58.5~e\,\mathrm{fm}^2$ are close to the corresponding data $-39.3(10)~e\,\mathrm{fm}^2$ and $63(2)~e\,\mathrm{fm}^2$. It is interesting to note that the magnitude of the $Q_s$ of the states with $K = 1/2$ and $K = 9/2$ presents different evolution trends  with the increase of $J$.  This behavior can be understood from the PRM, in which, the spectroscopic quadrupole moment $Q_s$ is simply given by the intrinsic quadrupole moment $Q_0$~\cite{Ring:1980}
\beq
\label{eq:Q_s_PRM}
Q^{\rm PRM}_s = Q_0\frac{3K^2 - J(J+1)}{(2J+3)(J+1)}.  
\eeq
For states with $K = 1/2$, the magnitude of $Q_s$ is dominated by the second term in the denominator of the above equation and therefore increases parabolically with $J$, as shown in Fig.~\ref{fig:Xe129_qs}(a). In contrast, for states with a large projection $K = 9/2$, the first term dominates at low $J$, while the second term gradually cancels its contribution as $J$ increases. At sufficiently high $J$, this cancellation can even reverse the sign of $Q_s$, as illustrated in Fig.~\ref{fig:Xe129_qs}(b). Moreover, one can see that the discrepancy between the MR-CDFT and PRM results in Fig.~\ref{fig:Xe129_qs}(a) increases with $J$, implying that the predominant shape of the state drifts toward a more deformed region as $J$ increases.

It is worth noting that in recent years the shape of $^{129}$Xe has been probed using relativistic heavy-ion collisions~\cite{STAR:2021mii,ATLAS:2022dov,Goldschmidt:2015kpa, Giacalone:2017dud,Giacalone:2019pca,Bally:2021qys,Zhao:2024lpc}.  These studies suggest the presence of triaxial deformation in $^{129}$Xe, and this finding is supported by the cranking shell model calculation~\cite{Huang:2016srl} and the MR-DFT calculation based on non-relativistic Skyrme forces~\cite{Bally:2021qys,Bally:2022rhf}. However, it was also pointed out that the energy surface is 
rather flat over a wide range of triaxial $\gamma$ values. It would be very interesting to know whether the triaxial energy minimum survivals with the presence of the octupole shape degree of freedom in the MR-DFT calculation in the future.

\subsection{\nuclide[199]{Hg}}

 The valence-space large-scale shell-model (LSSM) calculations have recently been carried out for \nuclide[199]{Hg} \cite{Sahoo:2024osx,Yanase:2020LSSM1}. In these studies, the valence space is defined by  the six orbitals ($0h_{9/2}, 1f_{7/2}, 1f_{5/2}, 2p_{3/2}, 2p_{1/2}, 0i_{13/}$). The effective charges $e_\pi = 1.5e$ and $e_\nu = 0.8e$, determined by reproducing the experimental $B(E2)$ values of \nuclide[200,202,204,206]{Hg}, were employed in the LSSM for nuclear electric observables. Recently, the DFT with AMP based on a Skyrme EDF was applied to investigate the magnetic and quadrupole moments of \nuclide[199]{Hg} without using effective $g$-factors or effective charges in the dipole or quadrupole operators \cite{Bonnard:2022xdr}. Here, we present the results of our MR-CDFT calculation for \nuclide[199]{Hg}.

\begin{figure}[]
 \centering
\includegraphics[width=8.5cm]{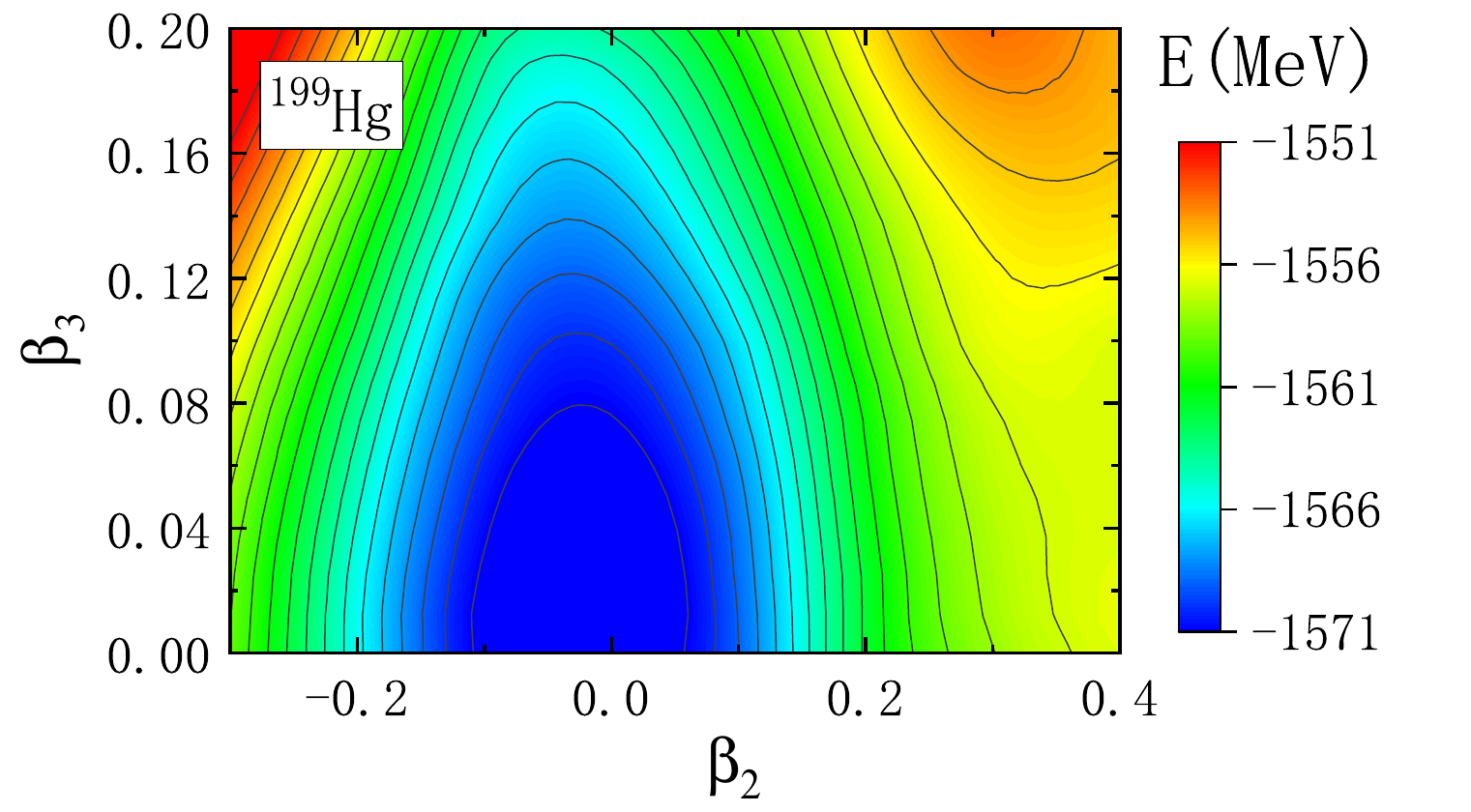}
\caption{Same as  the Fig.~\ref{fig:Xe129_MFE}, but for \nuclide[199]{Hg}.}
 \label{fig:hg199_MFE}
 \end{figure}

\begin{figure}[]
 \centering
\includegraphics[width=8.5cm]{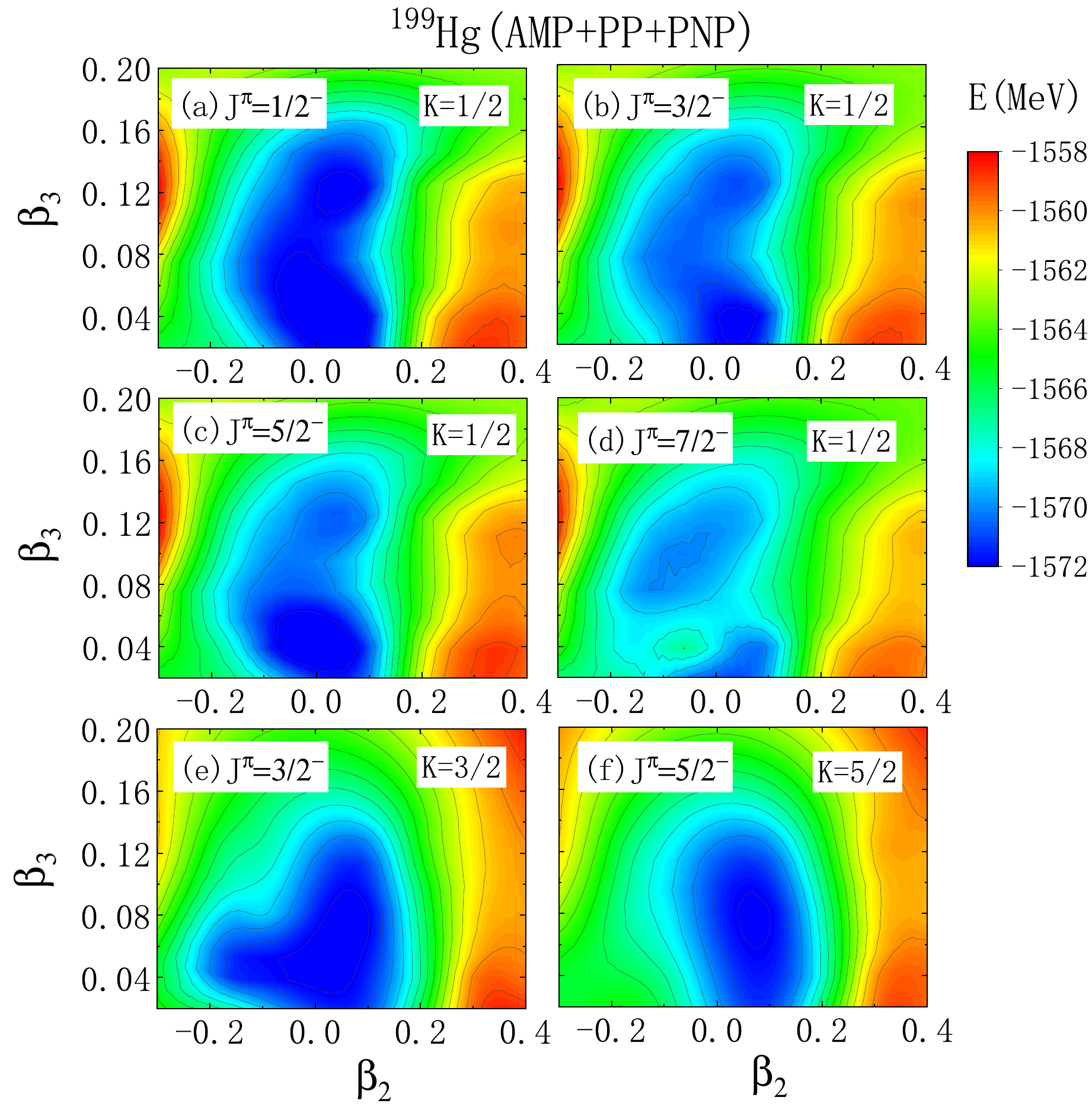}
\caption{Same as  the Fig.~\ref{fig:Xe129_PE}, but for \nuclide[199]{Hg} with the configurations of $K=1/2$  (a-d),  $3/2$ (e), and $5/2$ (f), respectively.
}
 \label{fig:Hg199_PE}
 \end{figure}

Figure~\ref{fig:hg199_MFE} shows the PES of \nuclide[199]{Hg} in the $(\beta_2, \beta_3)$ deformation plane. The energy-minimum state is located around the spherical shape, while exhibiting pronounced softness along both the quadrupole and octupole directions, implying significant shape fluctuations. Figure~\ref{fig:Hg199_PE} shows the PESs of \nuclide[199]{Hg} in $(\beta_2, \beta_3)$ deformation plane with projection onto $J^\pi$ and $(N,Z)$.  The configurations for the states $J^\pi$ are chosen as the one-quasiparticle states with $K = 1/2, 3/2$, and $5/2$, respectively. The energies of the corresponding global energy-minimum configurations for the  $J^\pi=1/2^-$,  $3/2^-$ and $5/2^-$ states with $K=J$ are $-172.333$, $-172.156$, and $-172.087$ MeV, respectively. For all three cases, the energy surfaces remain soft along the $\beta_3$ direction, indicating pronounced octupole vibrations compared with the mean-field results in Fig.~\ref{fig:hg199_MFE}. Furthermore, comparison of the PESs for the configurations with the same $K=1/2$ but with different spin parities ($J^\pi=1/2^-, 3/2^-, 5/2^-$) shows that increasing $J$ leaves $\beta_2$ nearly unchanged but slightly reduces $\beta_3$, while the minima become more concentrated, illustrating the stabilization of nuclear octupole shape with rotation~\cite{Yao:2015Ra224}.

     \begin{figure}[bt]
 \centering
\includegraphics[width=8.5cm]{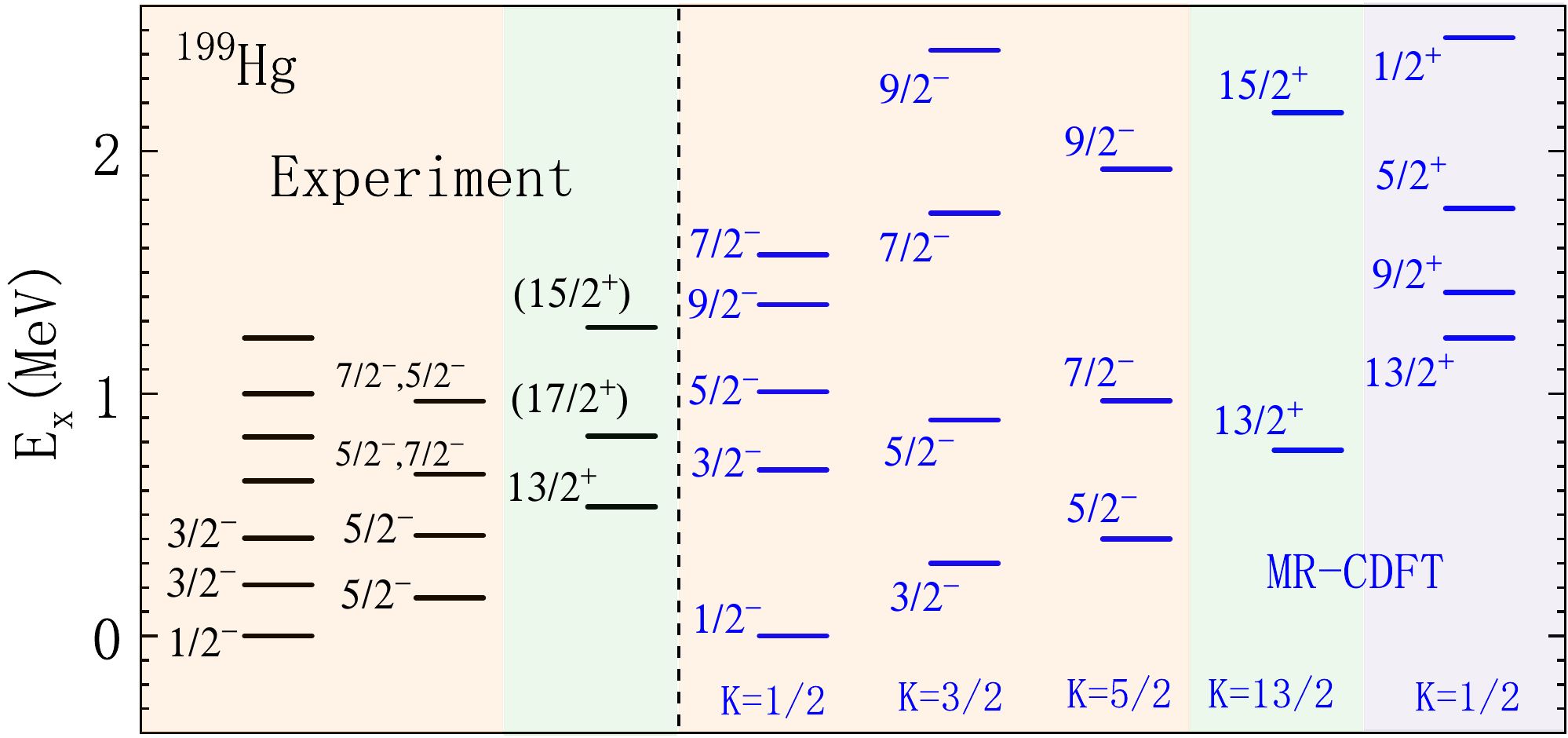}
\caption{Same as Fig.~\ref{fig:Xe129_spectra}, but for $^{199}$Hg.   }
 \label{fig:Hg199_spectra}
 \end{figure}

Figure~\ref{fig:Hg199_spectra} compares the low-lying states of \nuclide[199]{Hg} from MR-CDFT calculations with the available data. The states from the calculation are organized based on the $K$ quantum numbers of the configurations. It is seen that the main features of the low-energy structure are reasonably reproduced by the MR-CDFT, even though the spectrum is again systematically more stretched.  We note that there are several degenerate quasiparticle configurations with $K=1/2$, including the three ones with the unpaired neutron occupying the orbitals of $1/2[510], 1/2[501]$, and $1/2[521]$, splitting from the spherical $2f_{5/2}, 3p_{1/2}$ and $3p_{3/2}$ states, respectively. The mixing of different shapes results in the mixing of these configurations in the predicted ground state $1/2^-$. In other words, by incorporating quadrupole–octupole shape mixing, the MR-CDFT approach effectively accounts for the configuration mixing among multiple quasiparticle configurations with the same $K$, which is essential for reproducing the low-lying spectrum of \nuclide[199]{Hg}.

According to our calculation, the observed several low-lying $3/2^-, 5/2^-$ states can be grouped into the $K=3/2, 5/2$ bands, dominated by the configuration of $\nu 3/2[501]$ and $\nu 5/2[503]$, respectively,  splitting from the spherical $f_{5/2}$ states. Moreover, the sequences of states ($13/2^+, 17/2^+, 15/2^+$) on experiment can be classified as $K=13/2$ band, dominated by the configuration of $\nu 13/2[606]$. In addition,  we find the sequences of states ($13/2^+, 9/2^+, 5/2^+, 1/2^+$) from the MR-CDFT calculation and these states form a rotational band with $K=1/2$ built on top of the $13/2^+$ state which is dominated by the configuration of $\nu 1/2[660]$, splitting from the spherical $i_{13/2}$ state.  The energy ordering of these states follows the prediction of the PRM in the \textit{weak coupling limit}~\cite{Ring:1980}, which is realized for very small deformations and the unpaired neutron moves essentially on spherical shell model levels only slightly disturbed by shape vibrations. The lowest-energy state of this band is characterized by $J=j=13/2$. These states are to be confirmed experimentally.

\begin{figure}[]
 \centering
\includegraphics[width=8.5cm]{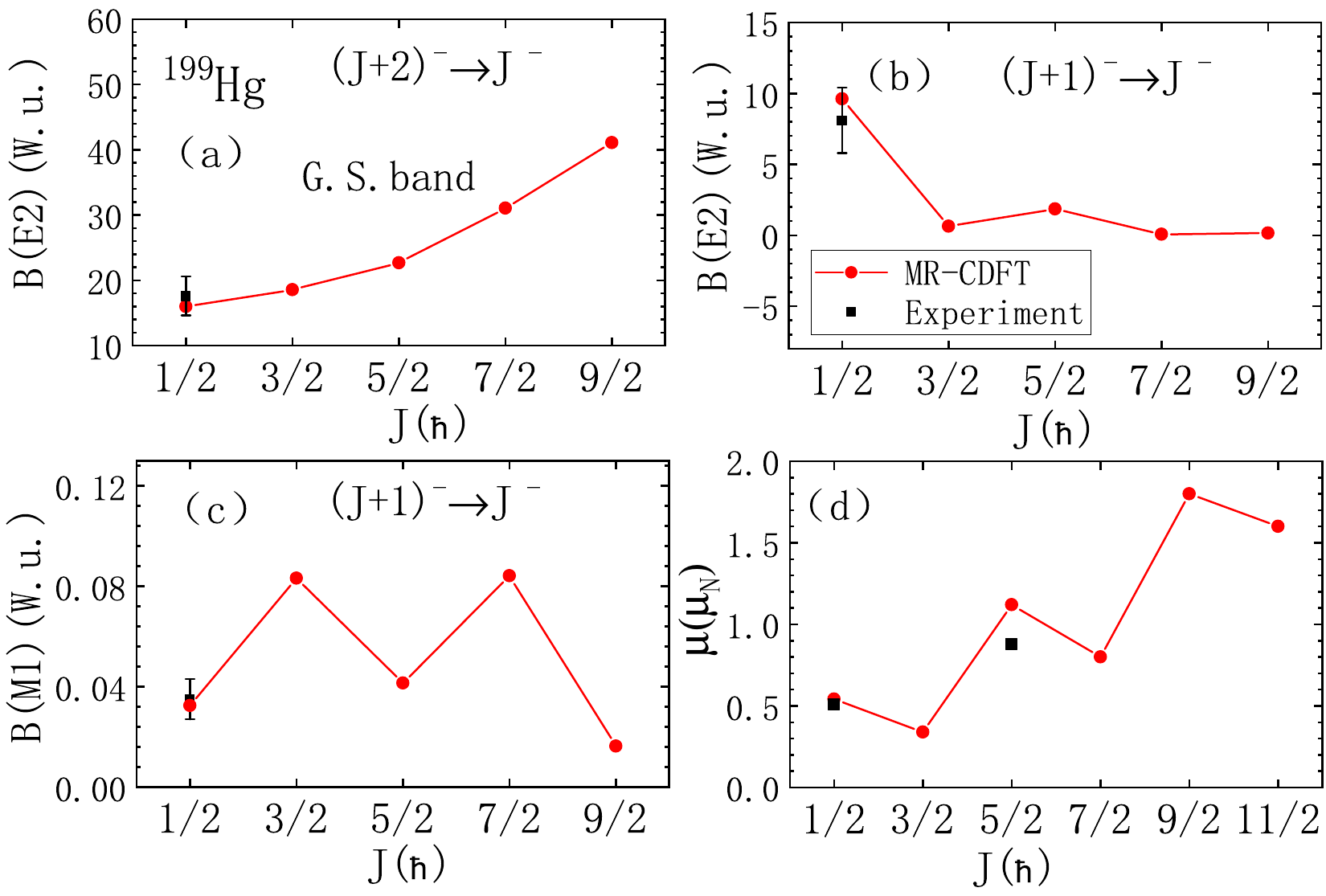}
\caption{ Same as Fig.~\ref{fig:Xe129_trans}, but for the ground-state band ($K=1/2$) in $^{199}$Hg.
}
 \label{fig:Hg199_EM}
 \end{figure}

\begin{table}[h!]
\centering
\tabcolsep=6pt
\caption{Spectroscopic quadrupole moments ($Q_s$) and magnetic dipole moments of the bandhead states with spin-parity $J^\pi = 1/2^-, 3/2^-, 5/2^-$, and $13/2^+$ in \nuclide[199]{Hg}, calculated with MR-CDFT and compared with experimental data~\cite{NNDC}. }
\begin{tabular}{lcccc}
  \hline \hline \\
   & \multicolumn{2}{c}{$Q_s$ (e~fm$^2$)} & \multicolumn{2}{c}{$\mu$ ($\mu_N$)} \\
   \cline{2-3} \cline{4-5}
  $J^{\pi}$ & MR-CDFT & Exp. & MR-CDFT & Exp. \\
  \hline
   $1/2^-$ & - & - & 0.54 & 0.506 \\ 
   $3/2^-$ & 47 & 50(12) & $-0.82$ & $-0.56(9)$ \\ 
   $5/2^-$ & 101 & 95(7) & 1.12 & 0.88(3) \\ 
   $13/2^+$ & 163 & 120(50) & $-1.26$ & $-1.014703(3)$ \\
  \hline \hline
\end{tabular}
\label{tab:Hg199_Qs_mu}
\end{table}

Figure~\ref{fig:Hg199_EM} shows the electric quadrupole transition strengths, magnetic dipole transition strengths, and magnetic moments of the ground-state band ($K=1/2$) in \nuclide[199]{Hg}. The calculated ground-state magnetic moment, $\mu(1/2^-) = 0.541\mu_N$, is in good agreement with the experimental value of $0.506\mu_N$. The variation of the transition strengths with angular momentum is similar to that observed in \nuclide[129]{Xe} (see Fig.~\ref{fig:Xe129_trans}). Both the magnetic dipole transition strengths, $B(M1)$, and the magnetic moments, $\mu$, exhibit a sawtooth-like increase with spin, which is again attributed to the Coriolis matrix elements~\cite{Bohr:1998v2}. Moreover, as shown in Table~\ref{tab:Hg199_Qs_mu}, the spectroscopic quadrupole moments ($Q_s$) and magnetic dipole moments ($\mu$) of the $J^\pi = 3/2^-$, $5/2^-$, and $13/2^+$ states in \nuclide[199]{Hg} are reasonably well reproduced. Considering that both $Q_s$ and $\mu$ are sensitive to configuration mixing, the successful reproduction of these quantities without introducing any free parameters further demonstrates the reliability of the MR-CDFT in describing the low-lying states of \nuclide[199]{Hg}.

 \subsection{\nuclide[225]{Ra}}

\begin{figure}[]
 \centering
\includegraphics[width=8.5cm]{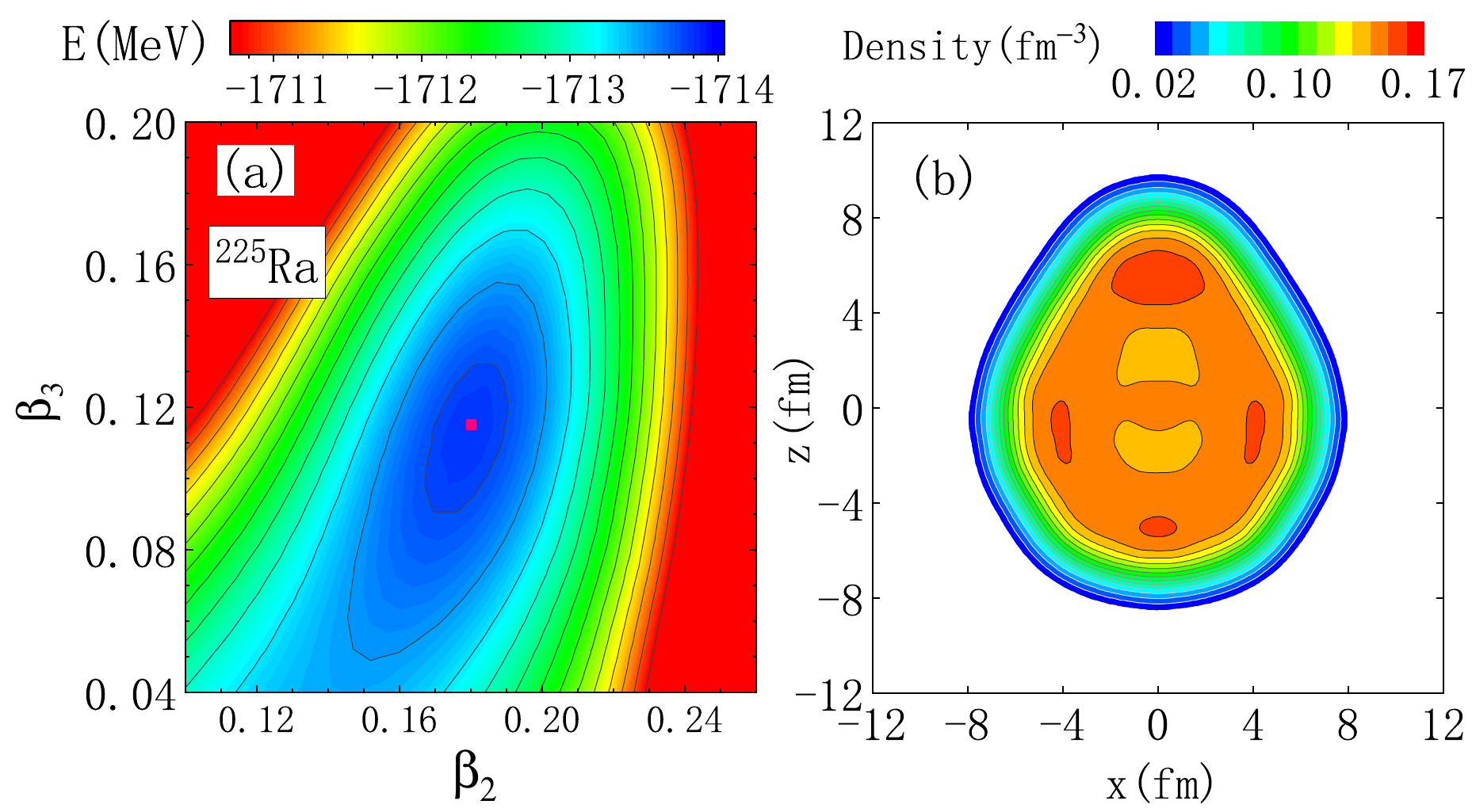}
\caption{
(a) Mean-field energy surface of \nuclide[225]{Ra} in the $(\beta_2,\beta_3)$ deformation plane.
(b) Total nucleon density of the energy-minimum state with $(\beta_2 = 0.18, \beta_3 = 0.11$) in the $x-z$ plane ($y=0$).
}
 \label{fig:Ra225_MFE}
 \end{figure}

The structural properties of \nuclide[225]{Ra} are of particular interest as this nucleus exhibits strong quadrupole-octupole correlations~\cite{Sheline:1983} and the corresponding atom may have an enhanced EDM~\cite{Engel:2013PPNP,Chupp:2019RMP}. This can be understood from the observed near degenerate parity doublets with spin parity of $1/2^\pm$ and the energy difference of only 55 keV~\cite{NNDC}, which leads to a two-to-three orders larger nuclear Schiff moment than \nuclide[199]{Hg}~\cite{Dobaczewski:2005PRL,Zhou:2025_letter}. The theoretical studies of the nuclear structure of  \nuclide[225]{Ra} remain scarce, except for a recent study with the core–quasiparticle coupling model~\cite{Zhao:2023}, primarily due to its structural complexity as a heavy nucleus with octupole degrees of freedom. In this subsection, we present the results of our MR-CDFT calculation for \nuclide[225]{Ra}. We note that our method has already been successfully applied to describe the energy spectrum and spectroscopic properties of \nuclide[224]{Ra}~\cite{Yao:2015Ra224}.

Figure~\ref{fig:Ra225_MFE} (a) displays the mean-field energy surface of \nuclide[225]{Ra}  in the $(\beta_2,\beta_3)$ deformation plane. A pronounced quadrupole-octupole deformed  minimum is found around  $(\beta_2 = 0.18$, $\beta_3 = 0.11)$, similar to  the case of \nuclide[224]{Ra}~\cite{Yao:2015Ra224}. The total nucleon density distribution of this mean-field state is shown in Fig.~\ref{fig:Ra225_MFE}, where a pear-shaped density profile is observed. The softness of the energy surface along the octupole direction suggests the potential important effect of octupole shape fluctuations. 

   \begin{figure}[]
 \centering
\includegraphics[width=8cm]{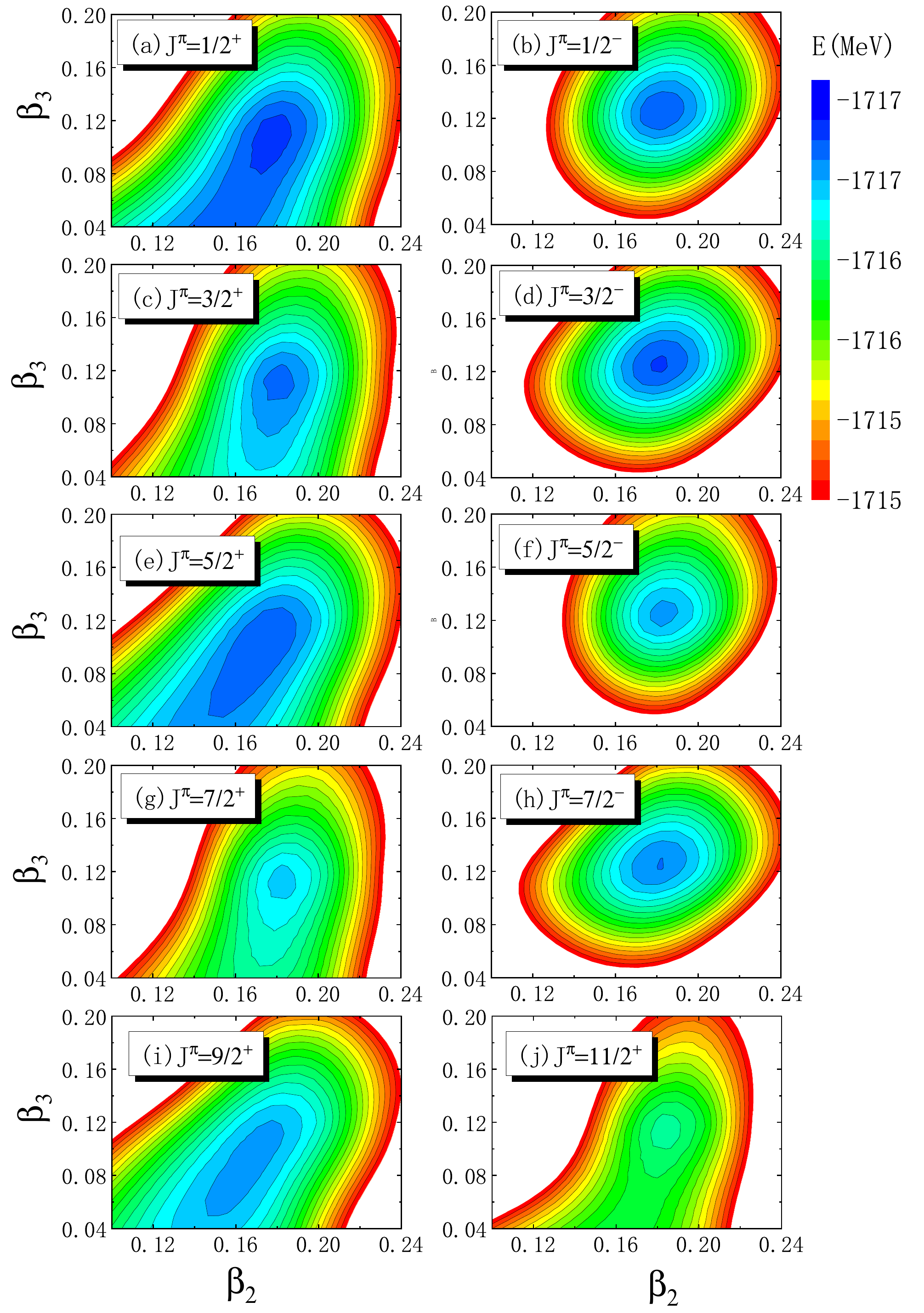}
\caption{The energy surfaces of quantum-number projected states for \nuclide[225]{Ra} in the $(\beta_2, \beta_3)$ deformation plane, corresponding to the quasiparticle configurations with $K=1/2$.
}
 \label{fig:Ra225_PE}
 \end{figure}

\begin{figure}[]
 \centering
\includegraphics[width=9cm]{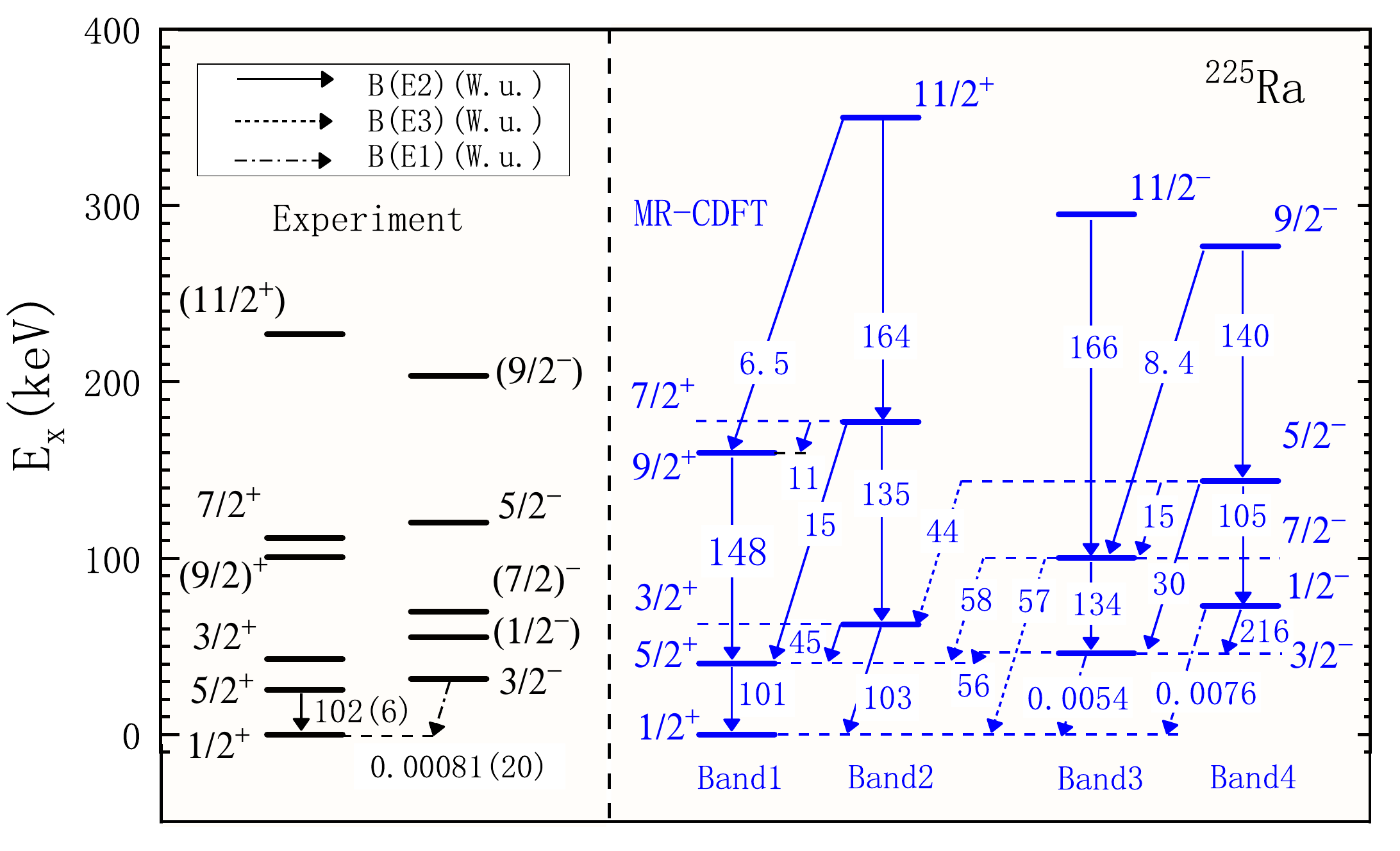}
\caption{Same as Fig.~\ref{fig:Xe129_spectra}, but for the states of $^{225}$Ra. The results of MR-CDFT calculation are for the configurations with $K=1/2$. The $B(E\lambda)$ values (in units of W.u.) are indicated near the arrows for the transitions. }
 \label{fig:Ra225_spectra}
 \end{figure}

Figure~\ref{fig:Ra225_PE} presents the PESs of quantum-number–projected states for \nuclide[225]{Ra} in the $(\beta_2,\beta_3)$ deformation plane. The lowest quasiparticle configuration with $K = 1/2$ is considered. It is shown that the PESs for $J^\pi = 1/2^+$, $5/2^+$, and $9/2^+$ are similar, those for $J^\pi = 3/2^+$, $7/2^+$, and $11/2^+$ are similar, and likewise, the PESs for $J^\pi = 1/2^-$ and $5/2^-$, as well as those for $J^\pi = 3/2^-$ and $7/2^-$, exhibit similar patterns. A closer inspection reveals a gradual evolution of the PESs with increasing angular momentum $J$. Specifically, Fig.~\ref{fig:Ra225_PE} shows that the quadrupole–octupole deformations of the predominant configurations for the $J^\pi = 1/2^+$, $5/2^+$, and $9/2^+$ states decrease with increasing $J$. This behavior is opposite to that found in \nuclide[224]{Ra}, where the dominant shapes of positive-parity states gradually evolve toward those of negative-parity states as $J$ increases. The different behaviors observed in odd-mass and even-even nuclei highlight the nontrivial polarization effects induced by the unpaired neutron. This phenomenon explains the main features of the  energy spectrum of  \nuclide[225]{Ra}. It is shown in Fig.~\ref{fig:Ra225_spectra}  that  the low-lying states can be approximately classified into four $\Delta J=2$ rotational bands, with the sequences of $(1/2^+, 5/2^+, 9/2^+, \cdots)$ for band 1, $(3/2^+, 7/2^+, 11/2^+, \cdots)$ for band 2, $(1/2^-, 5/2^-, 9/2^-, \cdots)$ for band 3 and $(3/2^-, 7/2^-, 11/2^-, \cdots)$ for band 4, respectively. These four bands are dominated by the admixture of the positive-parity configurations of $\nu1/2[631], \nu1/2[640]$, and negative-parity configuration of  $\nu1/2[501]$, due to the quadrupole-octupole deformations. The two bands having $\Delta J = 2$ each (even and odd values of $J + 1/2$) shifted against one another. It can be understood based on the PRM  in the \textit{strong coupling  limit}, where the angular momentum $\mathbf{j}$ of the valence particles is strongly coupled to the motion of the core, i.e., $\mathbf{j}$ precesses around the 3-axis. In this limit, the Coriolis interaction is small and its contribution only for the $K = 1/2$ bands can be considered in first-order perturbation theory. With this approximation, the energy of the $i$-state in the PRM is given by~\cite{Ring:1980},
\beqn
&&E^{\rm i, PRM}_{K=1/2}(J) \nonumber\\
&=& \mathcal{E}^i_{K=1/2} + \frac{1}{2\mathcal{J}}
\left\{ J(J+1) - \frac{1}{4} + a^i \left( J + \frac{1}{2} \right)(-)^{J + 1/2} \right\}, 
\eeqn
where $\mathcal{E}^i_{K=1/2}$ is the quasiparticle energy, $\mathcal{J}$ is the moment of inertia. If the so-called \textit{decoupling factor} $a^i$  is positive, the levels with odd values of $I + 1/2$ (band 1) are shifted downwards.   Interestingly, one observes an opposite behavior for the negative-parity  states, corresponding to the case with a negative value of $a^i$. A full understanding of this phenomenon requires a further dedicate study for the $a^i$, which is beyond the scope of this work.

\begin{figure}[]
 \centering
\includegraphics[width=6cm]{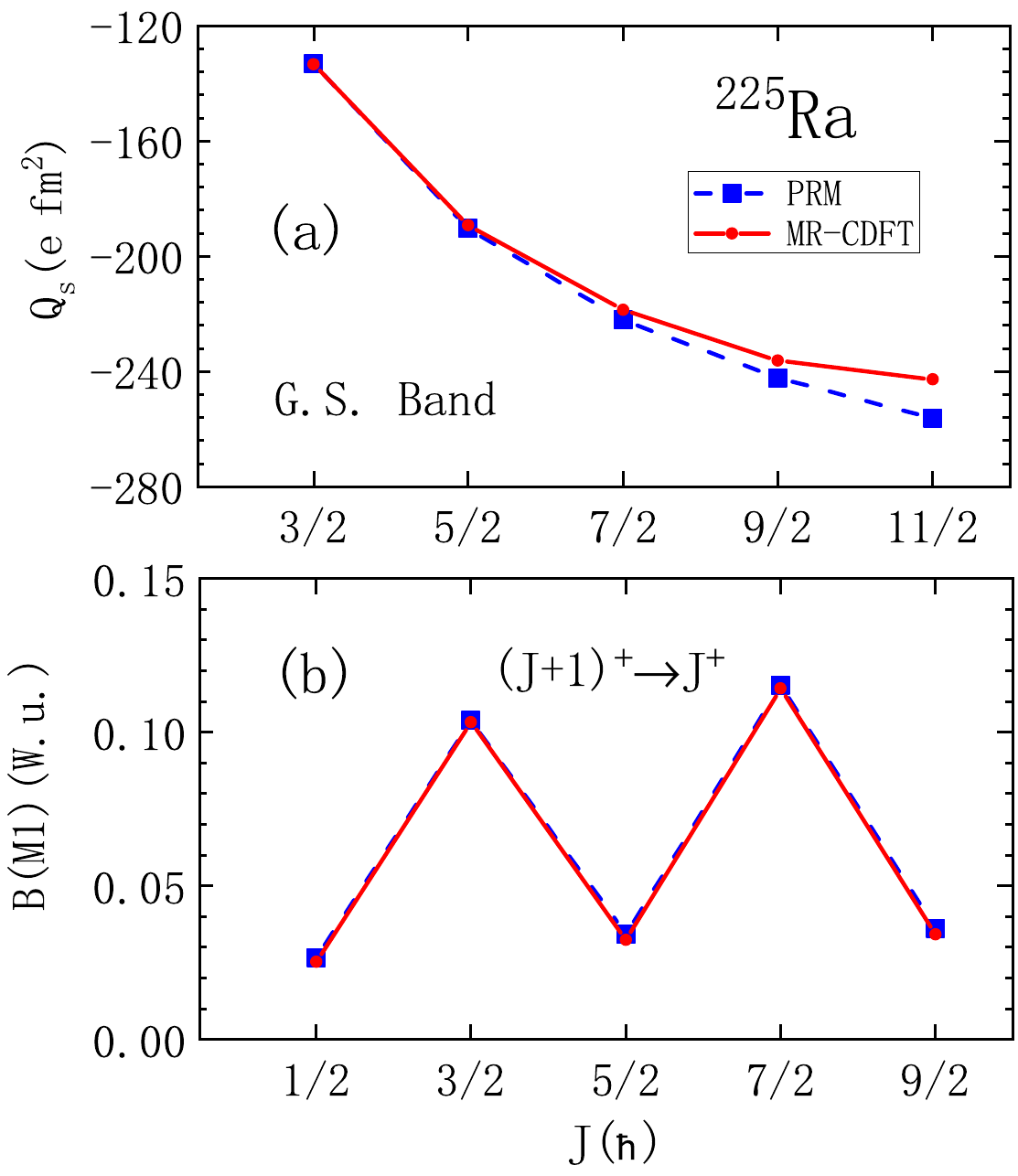}
\caption{ Spectroscopic quadrupole moments $Q_s$ (a) and magnetic dipole transition strengths $B(M1)$ (b) for the positive-parity states of \nuclide[225]{Ra} obtained from MR-CDFT calculations as a function of angular momentum, in comparison with the results of PRM calculations. See main text for details. 
}
 \label{fig:Ra225_QsM1}
 \end{figure}

 Figure~\ref{fig:Ra225_spectra} also provides the electric quadrupole transition strengths in \nuclide[225]{Ra}. It is seen that the transitions between $\Delta J=2$ states are significantly stronger than those for $\Delta J=1$ transitions. 
 The predicted $B(E2; 5/2^+ \rightarrow 1/2^+)=101$ (W.u.) is consistent with the data of 102(6) (W.u.). 
 Moreover, the excitation energy of the $1/2^-$ state is predicted to be 78 keV, slightly larger than the data 55 keV. 
 The predicted $E1$ transition strength $B(E1; 3/2^- \rightarrow 1/2^+)=0.0054$  (W.u.) is larger than the data 0.00081(20) (W.u.).   Figure~\ref{fig:Ra225_QsM1} shows the spectroscopic quadrupole moments $Q_s$ and magnetic dipole transition strengths $B(M1)$ of \nuclide[225]{Ra}. The change of the spectroscopic quadrupole moment with angular momentum follows the prediction of the PRM (\ref{eq:Q_s_PRM}), exhibiting a parabolic trend. The $B(M1)$ presents a staggering behavior with the increase of $J$, which can be understood from the formula of the PRM~\cite{Bohr:1998v2}
\begin{align}
&B^{\rm PRM}(M1; J_1 \rightarrow J_2) \nonumber\\
  &= \frac{3}{16\pi}\, \mu_N^2 (g_K - g_R)^2
     \left[ 1 + (-1)^{J_> + 1/2} b \right]
     \langle J_11/2; 10 | J_21/2 \rangle^2 .
\end{align}
where $J_>=\max(J_1, J_2)$. Since there is no data for the $B(M1)$ of \nuclide[225]{Ra}, 
the parameters $g_K - g_R$ and $b$ in the PRM are fitted to the results of MR-CDFT.
We find that $g_K - g_R=1.69$ and $b=0.53$. It is shown in Fig.~\ref{fig:Ra225_QsM1} that the main features of the $Q_s$ and $B(M1)$ for the positive-parity states of \nuclide[225]{Ra} from our MR-CDFT calculation can be rather well reproduced by the PRM, when the phenomenological parameters are optimized properly.

\begin{table}[tb]
\centering
\tabcolsep=20pt
\caption{Comparison of calculated and experimental magnetic moments (in $\mu_N$) for the low-lying states of \nuclide[225]{Ra}.}
\begin{tabular}{lcc}
\hline\hline
$J^\pi$ & MR-CDFT & Exp. \\
\hline
$1/2^+$ & $-0.785$ & $-0.7338(15)$ \\
$5/2^+$ & $0.082$ & $-$ \\ 
$3/2^+$ & $1.63$ & $-$ \\
$7/2^+$ & $2.82$ & $-$ \\ 
$1/2^-$ & $0.96$ & $-$ \\
$3/2^-$ & $-0.39$ & $-$ \\
\hline\hline
\end{tabular}
\label{tab:Ra225_moments}
\end{table}

Table~\ref{tab:Ra225_moments} lists the magnetic moments of the ground state $J^\pi = 1/2^+$ and several low-lying states $(1/2^-, 3/2^\pm, 5/2^+, 7/2^+)$ from the MR-CDFT calculations, in comparison with the available data. It can be seen that the predicted magnetic moment of the ground state $-0.785 \mu_N$ is  close to the data $-0.7338(15) \mu_N$. In addition, for the positive-parity states, the magnetic moment does not show a linear dependence on the angular momentum, different from \nuclide[25]{Mg}~\cite{Zhou:2024PRC} and \nuclide[229]{Th}, as shown later. This difference can be understood from the fact that the quadrupole deformation of \nuclide[225]{Ra} is smaller than those of  \nuclide[25]{Mg} and \nuclide[229]{Th}. Moreover, we predict the magnetic moment of the $1/2^-$ state in \nuclide[225]{Ra} to be $0.96 \mu_N$, which is significantly different from that of the ground state with the same angular momentum and awaits confirmation from future experimental measurements.

\subsection{\nuclide[229]{Th} and \nuclide[229]{Pa}}

The nucleus \nuclide[229]{Th}, which has an isomeric state located approximately 8.4 eV above the ground state, has recently attracted tremendous interest due to its potential applications in nuclear clocks and precision tests of fundamental symmetries~\cite{Beck:2007zza,Beck:2009,Yoshinaga:2018PTSM4,Seiferle:2019fbe,E.Peik_2003,Peik:2015ota,Campbell:2012zzb,Flambaum:2006ak,Berengut:2009zz,Rellergert:2010zz,Fadeev:2020bgt,Tkalya:2010df}. Several nuclear models have been applied to describe the properties of the isomeric state~\cite{Muller:2018krl,Tkalya:2015xia,Ruchowska:2006dq,He:2007zzc,He:2008zzf,Litvinova:2009vp}, including its energy spectrum and electromagnetic characteristics. These include the hybrid collective quadrupole–octupole model (CQOM) and the deformed shell model (DSM)~\cite{Minkov:2021ovq,Minkov:2017kju,Minkov:2018qqm}, where a coupling mechanism between quadrupole–octupole deformation and single-particle motion was introduced to explain the origin of the extremely close energy levels. The model emphasizes the decisive role of octupole deformation in the formation of the isomeric state and successfully reproduces experimental observables with several adjustable parameters. More recently, the projected shell model (PSM)~\cite{Chen:2025rlz} has provided a reasonable description of the low-lying states of \nuclide[229]{Th}, with some adjustable parameters in nuclear Hamiltonian and effective charges for neutrons and protons. Therefore, a more microscopic and self-consistent method is still lacking for the study of low-lying and isomeric states in \nuclide[229]{Th}. The Skyrme Hartree–Fock+BCS approach has also been applied to investigate \nuclide[229]{Th}~\cite{Minkov:2024wna}. These studies suggest that \nuclide[229]{Th} exhibits pronounced quadrupole–octupole deformation, possibly even larger than that of \nuclide[225]{Ra}, indicating that \nuclide[229]{Th} may possess an even more enhanced Schiff moment.

Compared to \nuclide[229]{Th}, the nucleus \nuclide[229]{Pa} differs only by converting one neutron into one proton, and both nuclei belong to the actinide region. Therefore, it is expected that the two nuclei share similar low-energy structural properties. However, the available data on \nuclide[229]{Pa} are much scarcer than those for \nuclide[229]{Th}, and even its ground-state configuration has not yet been firmly established. \nuclide[229]{Pa} is of particular interest because it is one of the few nuclei in this region that exhibit $K = 1/2$ bands with different signatures. The low-lying states of \nuclide[229]{Pa} have been studied using a pairing-plus-multipole interaction Hamiltonian, which predicts a strong octupole deformation in the ground state and a nearly degenerate parity doublet $5/2^\pm$~\cite{Chasman:1980}. This parity-doublet structure, with a splitting energy of $0.22 \pm 0.05$~keV, was identified through studies of $\gamma$-ray and conversion-electron transitions following the $^{229}$U electron-capture decay~\cite{Ahmad:1982}, but it was not observed in the $^{231}$Pa($p,t$)$^{229}$Pa and $^{230}$Pa($p,2n\gamma$)$^{229}$Pa reactions~\cite{Grafen:1991,Levon:1994NPA}. If the parity doublet $5/2^\pm$ indeed exists, the Schiff moment of \nuclide[229]{Pa} would be enhanced by several orders of magnitude~\cite{Haxton:1983,Auerbach:1989}. Clarifying this controversy requires further dedicated experimental investigations, although such measurements are extremely challenging. In this subsection, we present a comprehensive study of the low-energy structural properties of both \nuclide[229]{Th} and \nuclide[229]{Pa} using our MR-CDFT framework. The results for these two nuclei are discussed side by side.

\begin{figure}[]
 \centering
\includegraphics[width=8.5cm]{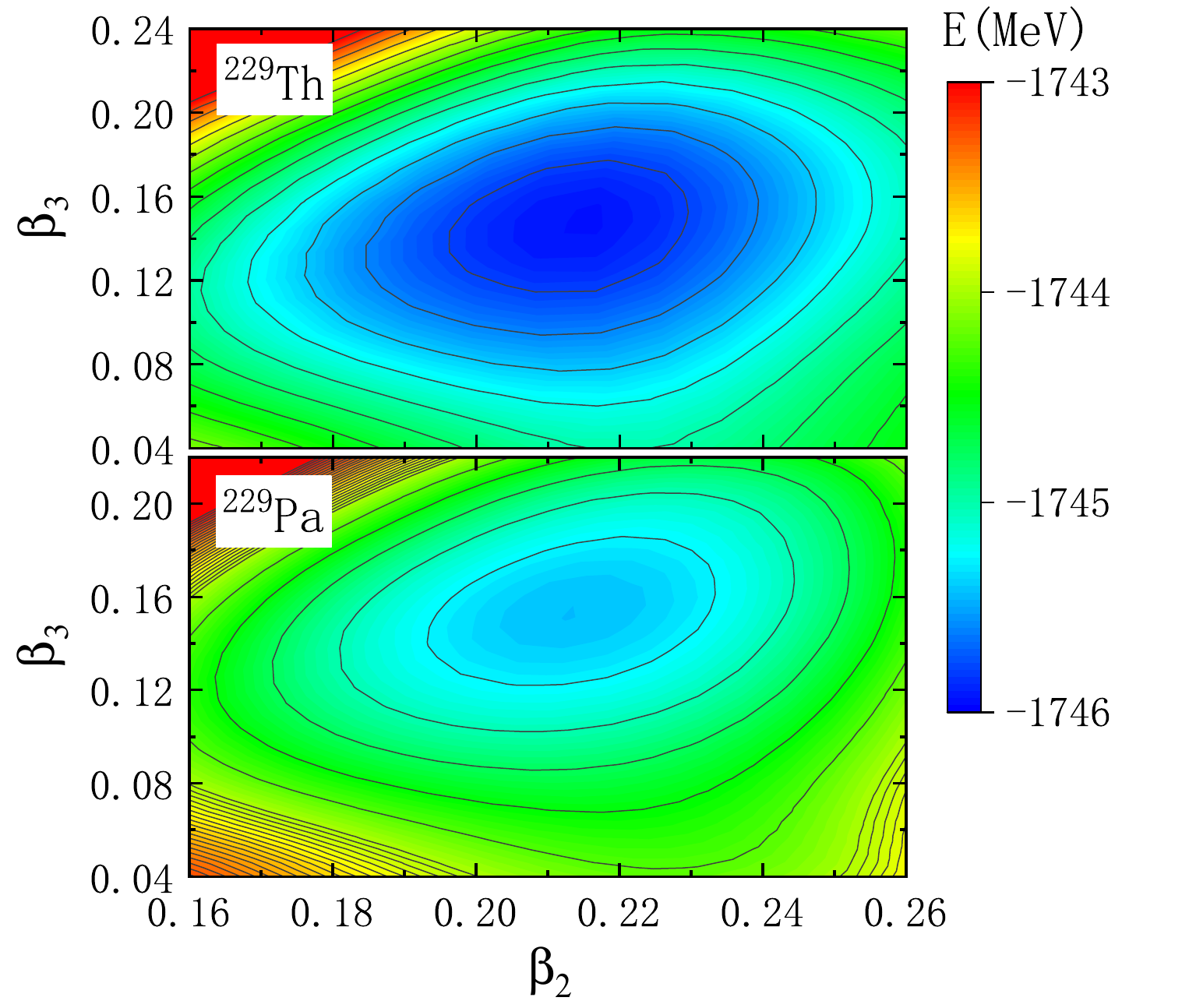}
\caption{The mean-field energy surfaces for \nuclide[229]{Th} and \nuclide[229]{Pa}. The energy difference between neighboring two contour lines is 0.2 MeV.}
 \label{fig:Th229_Pa229_MFE}
 \end{figure}

   \begin{figure}[]
 \centering
\includegraphics[width=\columnwidth]{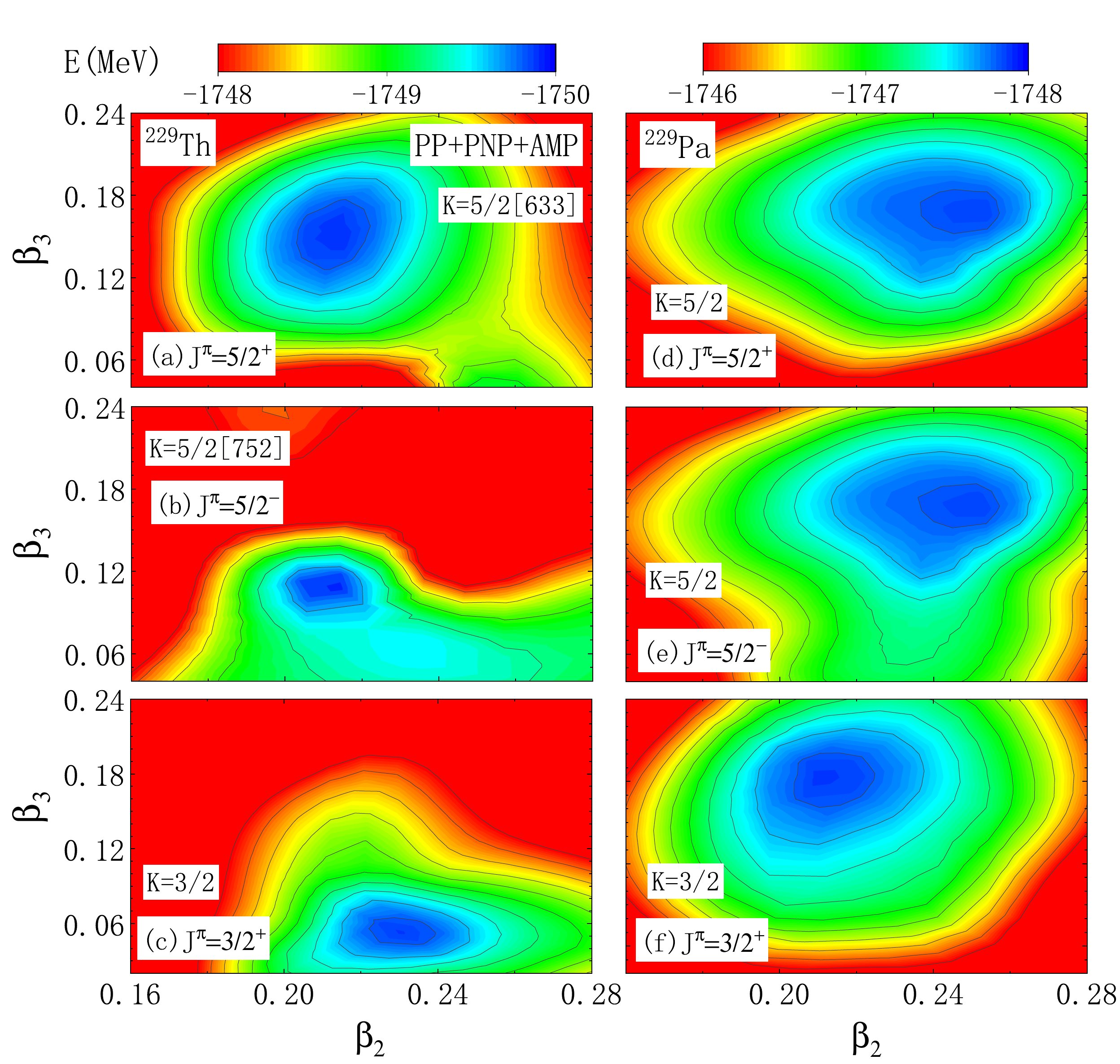}  
\caption{Comparison of the PESs for the projected states of $^{229}$Th (a-c) and $^{229}$Pa (d-f) with different spin parities.
For $^{229}$Th, the main quasiparticle configurations contributing to panels (a) and (b) are $\nu5/2[633]$ and $\nu5/2[752]$, respectively.
 }
 \label{fig:Th229_Pa229_PE} 
\end{figure}

 \begin{figure*}[bt]
 \centering
\includegraphics[width=0.6\paperwidth]{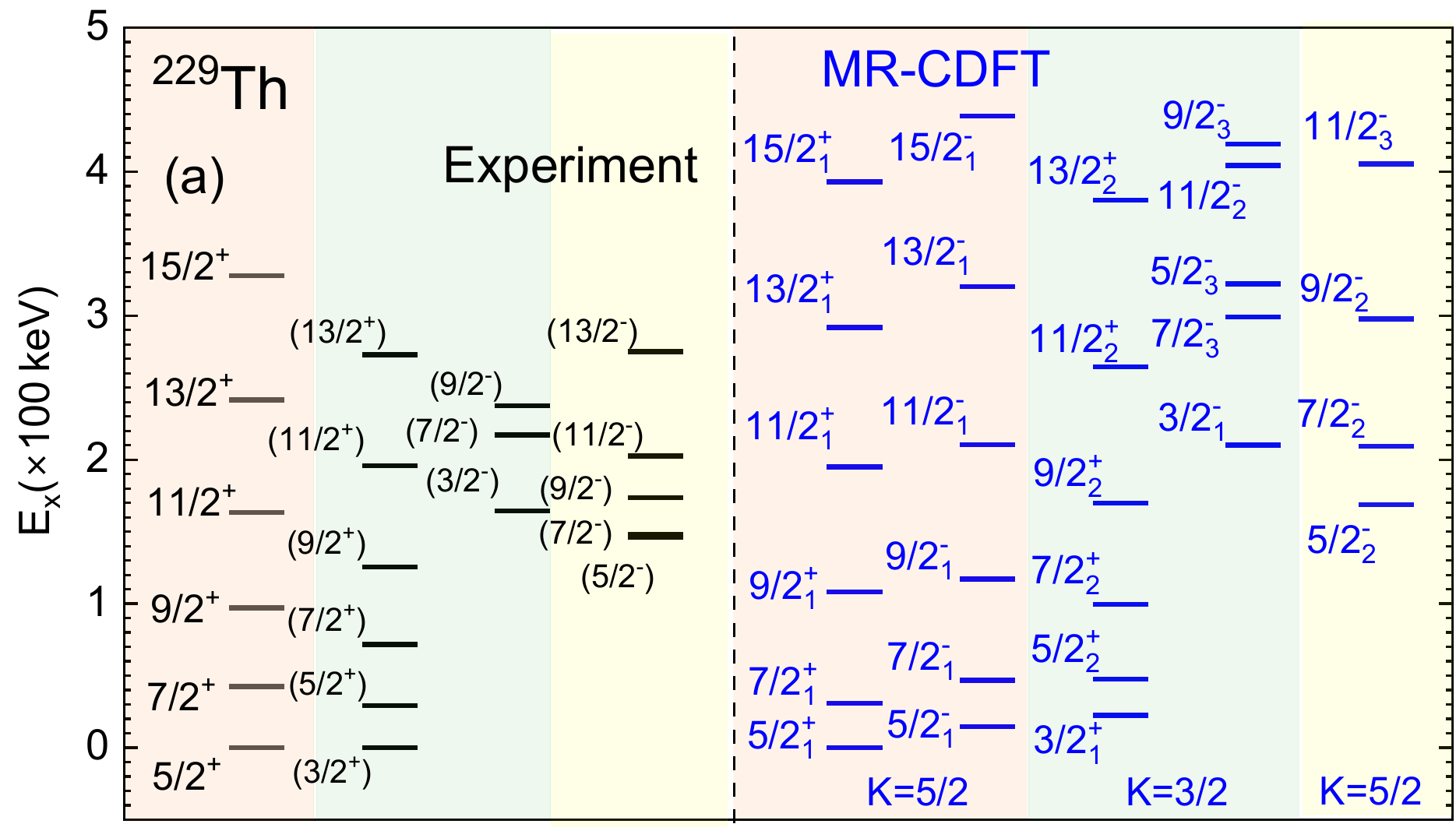} 
\includegraphics[width=0.6\paperwidth]{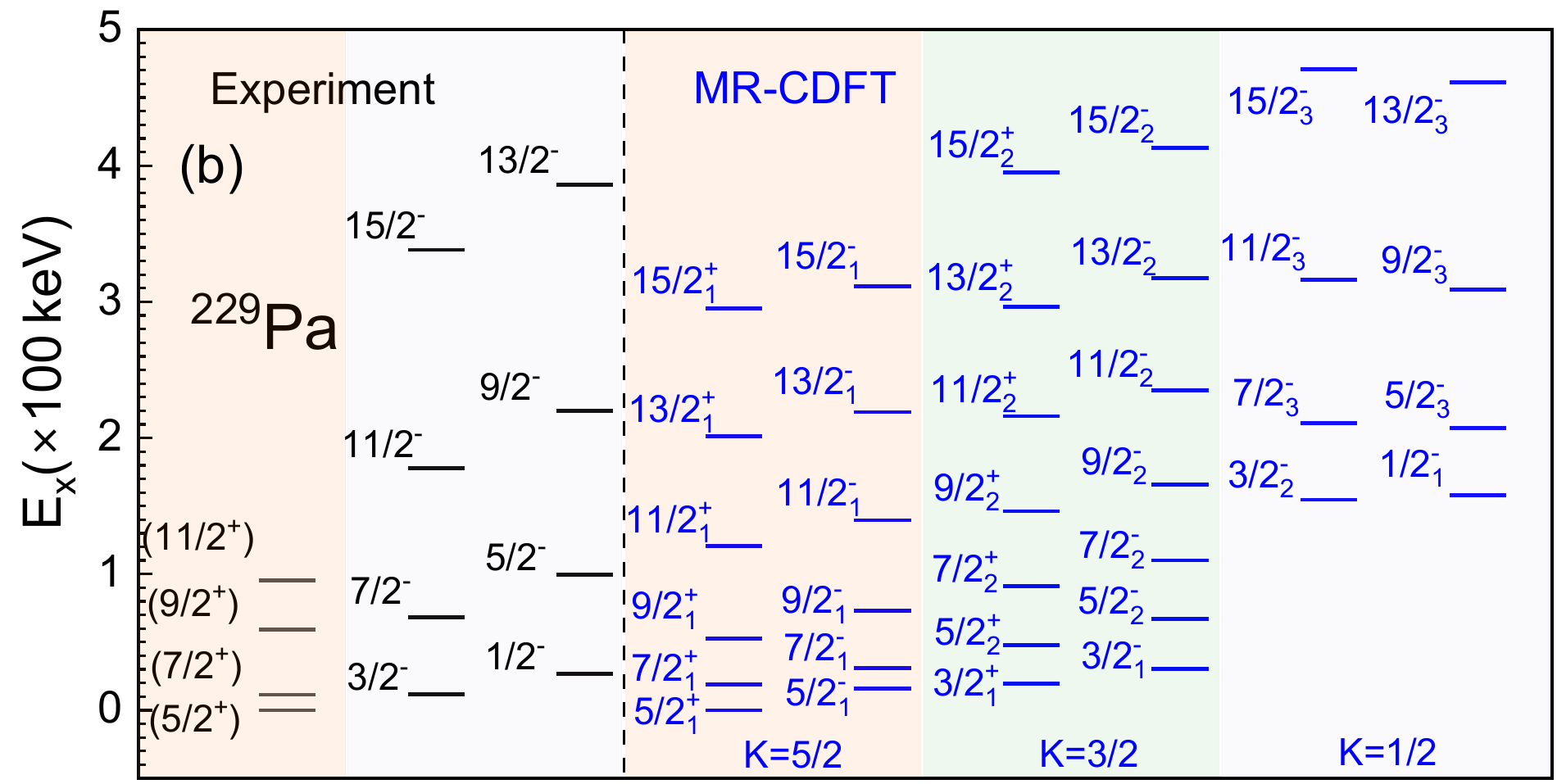} 
\caption{The energy spectra of low-lying states in $^{229}$Th (a) and $^{229}$Pa (b). }
 \label{fig:Th229_Pa229_Level}
 \end{figure*}

Figure~\ref{fig:Th229_Pa229_MFE} shows the mean-field PESs of $^{229}$Th and $^{229}$Pa in the $(\beta_2,\beta_3)$ deformation plane. The pronounced global energy minimum of  $^{229}$Th is located at $(\beta_2,\beta_3)=(0.22,0.15)$, significantly larger than the deformations of the energy minimum state of $^{225}$Ra, as shown in Fig.~\ref{fig:Ra225_MFE}. It is interesting to note that the PES around the energy minimum is soft along the quadrupole deformation. It is seen from Fig.~\ref{fig:Th229_Pa229_MFE}(b) that the PES of $^{229}$Pa
is  nearly identical to that of $^{229}$Th, except that the energies of all states are overall increased by about 0.5 MeV. The PESs of the projected states for  $^{229}$Th and $^{229}$Pa with different spin parities  are displayed in Fig.\ref{fig:Th229_Pa229_PE}. 
 We find that once the octupole deformation of the configuration $\beta_3> 0.14$, one obtains a nearly degenerate parity-doublet states, because the overlap of the space-reversion operator $\hat P$ between the wave function of this configuration vanishes~\cite{Yao:2015Ra224}. For instance, the lowest-energy state with $J^\pi = 5/2^- $ is nearly degenerate with the that with $J^\pi = 5/2^+$. The global energy minimum of the configuration $\nu 5/2[633]$, projected onto $J^\pi = 5/2^+$ is located around  $(\beta_2, \beta_3) = (0.22, 0.16)$ with an energy of $-1749.808~\mathrm{MeV}$. The configuration $\nu 5/2[752]$, close in energy to $\nu 5/2[633]$, also exhibits a pronounced minimum after projection, as shown in Fig.~\ref{fig:Th229_Pa229_PE}(b), with the lowest projected energy corresponding to $J^\pi = 5/2^-$ at $(\beta_2, \beta_3) = (0.21, 0.11)$. For the $J^\pi = 3/2^+$ state, energy minimum is found at $(\beta_2, \beta_3) = (0.23, 0.05)$ with an energy of $-1749.768~\mathrm{MeV}$. 
   A similar phenomenon is found in $^{229}$Pa. Quantitatively, we find that the global energy minima for the $J^\pi=5/2^\pm$ states of $^{229}$Pa are located around $(\beta_2,\beta_3)=(0.25,0.16)$ with approximately the same energy $-1747.798$ MeV, while that of $J^\pi=3/2^+$ is located at $(0.21,0.16)$ with the energy $-1747.762$.  In addition, we find that the configuration with $K=1/2$ corresponds to an energy-minimum solution for the $3/2^{-}$ state around $(0.23,\,0.16)$, with a calculated energy of $-1747.611$ MeV. This result is not shown in the Fig.~\ref{fig:Th229_Pa229_PE}. The mixing of all the deformed configurations in Fig.\ref{fig:Th229_Pa229_PE} gives the final physical state in the MR-CDFT.

Figure~\ref{fig:Th229_Pa229_Level} displays the energy spectra of $^{229}$Th and $^{229}$Pa. It should be noted that, unlike the calculations for \nuclide[129]{Xe}, \nuclide[199]{Hg}, and \nuclide[225]{Ra}, we also consider the mixing of quasiparticle configurations with different $K$ quantum numbers for these two nuclei and find that the $K$-mixing effect is generally minor (less than 3\%) for both the energy spectra and intraband transitions. Therefore, all states are labeled by the $K$ quantum numbers of the predominant quasiparticle configurations. The main features of the low-lying states of both nuclei are reproduced reasonably well.

Quantitatively, we find that the binding energies of the  ground states of $^{229}$Th and $^{229}$Pa with spin-parity $5/2^+$, calculated using the MR-CDFT, are 1750.366~MeV and 1749.093~MeV, respectively, compared with the experimental values of 1748.335~MeV and 1747.241~MeV. In addition, our main findings for $^{229}$Th are summarized as follows:
\begin{itemize}
    \item The experimental data on the energy spectrum indicate the existence of a ground-state rotational band (built on the $5/2^+$ state) and an excited rotational band (built on the isomeric $3/2^+$ state), whose excitation energies approximately follow the $J(J+1)$ rule. This feature is well reproduced by the MR-CDFT calculations, which reveal that these two bands can be mainly characterized as $K = 5/2$ and $K = 3/2$ bands, respectively. The predominant configurations for the $K=5/2$ bands are the admixture of components $\nu 5/2[633]$, $\nu 5/2[503]$ and $\nu 5/2[752]$, while that for the rotational band on top of the isomeric state is $\nu 3/2[631]$. The excitation energy of the isomeric $3/2^+$ state is predicted to be about 23~keV, whereas the experimental value is 8.4~eV. Reproducing such an atomic-scale excitation energy is beyond the reach of any current nuclear model.

    \item As shown in Fig.~\ref{fig:Th229_Pa229_PE}(a), the quadrupole–octupole deformations of the energy-minimum state with $J^\pi = 5/2^+$ are quite large. In this case, we predict a nearly degenerate negative-parity band as the partner of the ground-state band in \nuclide[229]{Th}, even though these states have not yet been observed experimentally. The excitation energy of the bandhead $5/2^-$ state is about 15~keV. If this state exists, the Schiff moment is expected to be enhanced by approximately one order of magnitude~\cite{Zhou:2025_letter}.
    
\item  In contrast to the ground-state rotational band which exhibits strong quadrupole–octupole deformation, the excited rotational band $(K=3/2)$ is characterized by strong quadrupole deformation but weak octupole deformation. Thus, it is likely that the nearly degenerate negative-parity partner states of the ground-state rotational band are missing in the experimental observations. The sequence of states $(3/2^-, 7/2^-, 9/2^-)$ should be classified as the negative-parity partners of the excited rotational band.

    \item The rightmost column of Fig.~\ref{fig:Th229_Pa229_Level}(a) shows another sequence of states, $(5/2^-,7/2^-,9/2^-,\ldots)$. The experimental excitation energy of the $5/2^-$ state is 146~keV. In our MR-CDFT calculation, we find a rotational band at similar energies based on the predominant configuration $\nu5/2[752]$, consistent with the findings of the Skyrme HF+BCS calculation~\cite{Minkov:2024wna}. This band has a slightly different predominant configuration from that of the ground-state band, as discussed before. 
    
\end{itemize}

For comparison, we highlight the main findings from the energy spectrum of $^{229}$Pa:
\begin{itemize}
    \item According to our MR-CDFT calculations, the predominant configurations for the ground state $5/2^+$ are the negative-parity configuration $\pi 5/2[523]$ and the positive-parity configuration $\pi 5/2[642]$, mixed by octupole deformation.   Based on these configurations with $K = 5/2$, we identify a $\Delta J = 1$ ground-state rotational band. Owing to the strong octupole deformation of the configuration, we obtain an negative-parity partner band with the excitation energy of the bandhead $5/2^-$ state  $E_x(5/2^-)=16~\mathrm{keV}$, which is very close to the excitation energy $15~\mathrm{keV}$ of a similar state in $^{229}$Th. It confirms the prediction of the earlier theoretical study based on the pairing-plus-multipole interaction~\cite{Chasman:1980} and is supported by the measurement of $\gamma$ and conversion-electron transitions in the $^{229}$U electron-capture decay.~\cite{Ahmad:1982}, even though the negative-parity partner band was not found in the later measurements\cite{Grafen:1991}. Moreover, we note that the observed sequence of states ($5/2^+$, $7/2^+$, $9/2^+$, $11/2^+$) does not behave as a typical rotational band, which requires further investigation. 
 \item The observed negative-parity states can be classified into two rotational bands with different signatures, both based on the predominant configuration $\pi 1/2[530]$. These findings are consistent with those of the previous study~\cite{Levon:1994NPA}. The energy shift between these two signature bands is attributed to the decoupling parameter arising from the Coriolis term~\cite{Ring:1980}. Compared with the experimental data, the energies of the two signature bands are overall overestimated by the MR-CDFT, whereas the main features of the rotational bands are well reproduced.

 \item In the MR-CDFT results, we also identify another set of parity-doublet bands (labeled by $K = 3/2$) built on the admixture of the $\pi 3/2[651]$ and $\pi 3/2[521]$ configurations caused by octupole deformation. The excitation energy of the bandhead $3/2^+$ state is $E_x(3/2^+)=32~\mathrm{keV}$, which is comparable to that of the isomeric state in $^{229}$Th with $23~\mathrm{keV}$. These $K=3/2$ parity-doublet bands have not been discussed before.

\end{itemize}

Comparing the energy spectra of $^{229}$Th and $^{229}$Pa obtained from the MR-CDFT calculations in Fig.\ref{fig:Th229_Pa229_Level}, one finds that the low-lying states of these two nuclei are overall similar, even though their predominant configurations are different. The main difference in the low-energy structure lies in the energy shift between the parity-doublet bands with $K = 3/2$. These two bands are predicted to be nearly degenerate in $^{229}$Pa because of the existence of stronger octupole correlations than that in $^{229}$Th, as shown in Fig.~\ref{fig:Th229_Pa229_PE}(c) and (f).

    \begin{figure}[]
 \centering
\includegraphics[width=8.5cm]{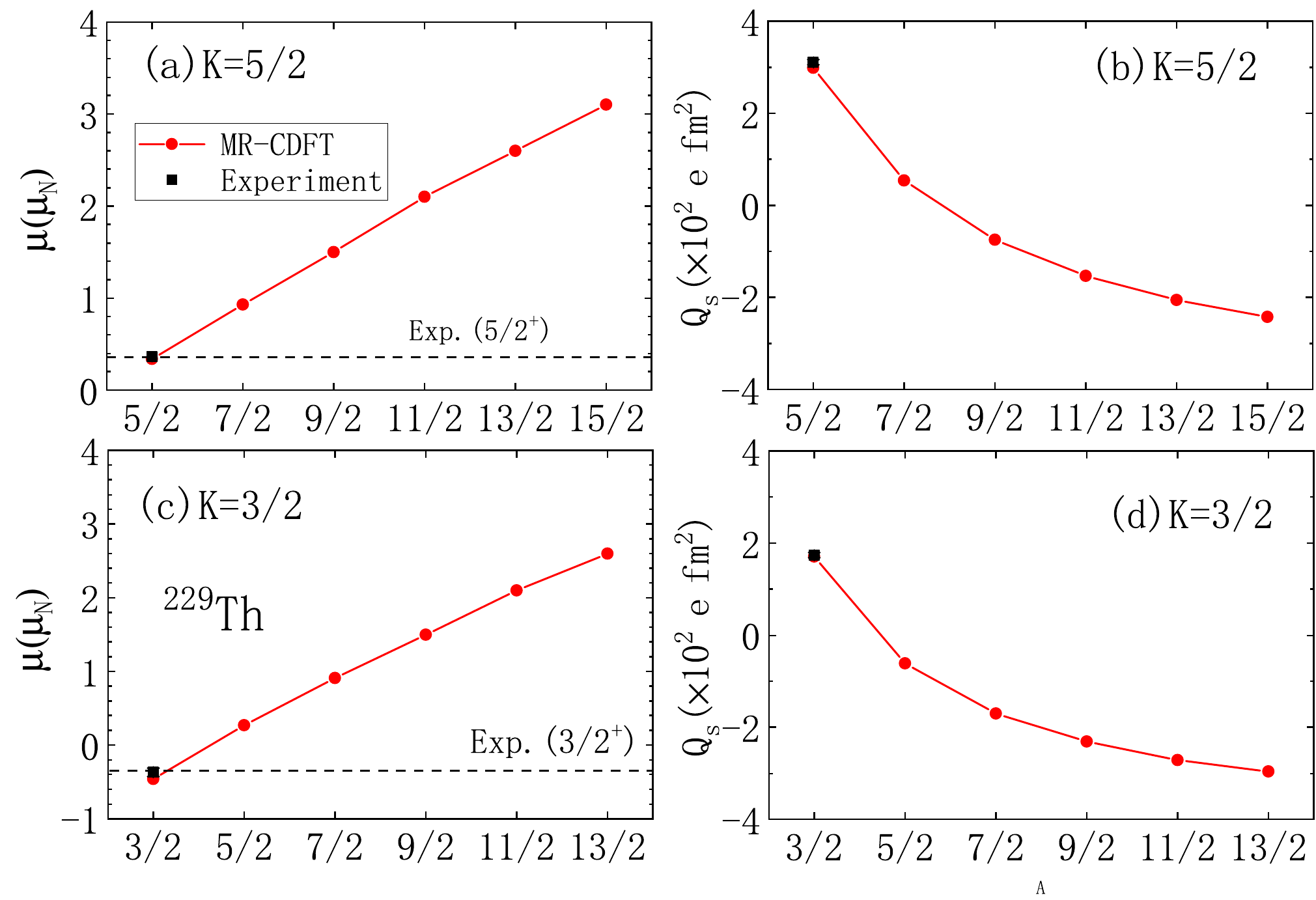}
\caption{ 
Magnetic moments $\mu$ and spectroscopic quadrupole moments $Q_s$ of the ground-state band ($K=5/2$) (a,b) and excited band ($K=3/2$) (c,d) in \nuclide[229]{Th} as functions of angular momentum, calculated with the MR-CDFT. The data is from Refs.~\cite{Peik:2020cwm,Safronova:2013xla,Zitzer:2025dti,Thielking:2017qet,Muller:2018krl}.
}
 \label{fig:Th229_mu_Qs}
 \end{figure}

 \begin{figure}[]
 \centering
\includegraphics[width=6cm]{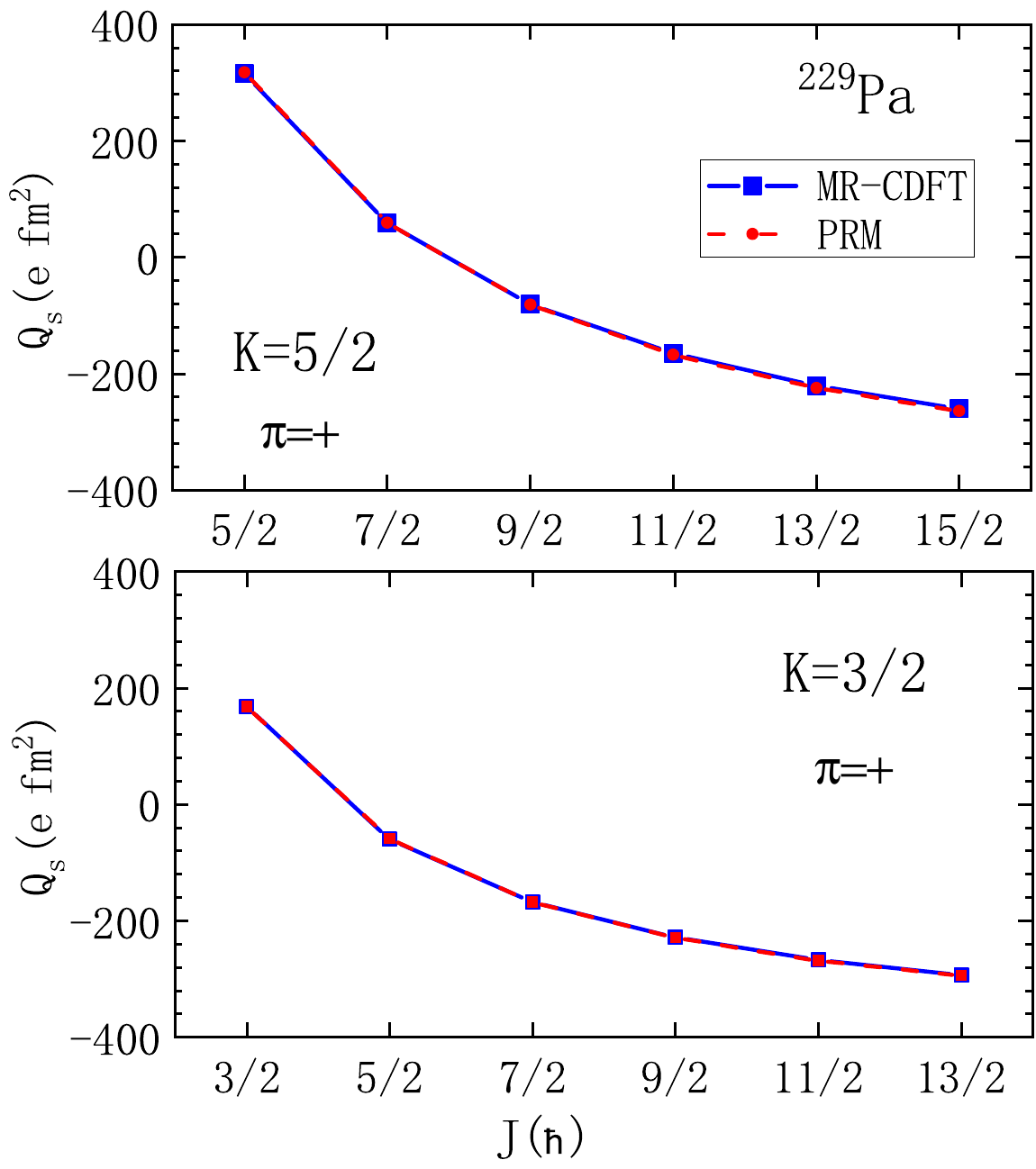}
\caption{ 
Spectroscopic quadrupole moments of the positive-parity bands in $^{229}$Pa as a function of angular momentum, calculated with MR-CDFT for quasiparticle orbitals $K=5/2$ and $K=3/2$. The corresponding results from the PRM are also provided for comparison.
}
 \label{fig:Pa229_Qs}
 \end{figure}

\begin{table}[h!]
\centering
\tabcolsep=5pt
\caption{
Magnetic moments $(\mu)$, spectroscopic quadrupole moments $(Q_s)$, and excitation energies $E_x$ of several low-lying states in $^{229}$Pa, including $J^\pi_\alpha = 5/2_1^{\pm}, 3/2_1^{\pm}, 1/2_1^{-}$ and $3/2_2^{-}$, calculated with the MR-CDFT.
 }
\begin{tabular}{lcccccc}
  \hline \hline \\
  $J^{\pi}$ & $5/2^+_1$ & $5/2_1^-$ & $3/2^+_1$ & $3/2^-_1$ & $1/2^-_1$ & $3/2^-_2$\\
  \hline
      $E_x$(keV)    & 0 & 16    &  19    & 31   & 158    & 155 \\ 
   $\mu(\mu_N)$ & $2.26$    & 2.21 & 2.02 & 2.03 &0.19& $1.57$ \\ 
   $Q_s$(e~fm$^2$)    & $319$ & 318    &  166    & 165   & -    & 177 \\ 
  \hline \hline
\end{tabular}
\label{tab:Th229}
\end{table}

Figure~\ref{fig:Th229_mu_Qs} shows the magnetic moments and spectroscopic quadrupole moments of the ground-state band ($K = 5/2$) and the isomeric band ($K = 3/2$) in $^{229}$Th as functions of the angular momentum $J$. It is seen that the predicted magnetic moments exhibit an approximately linear dependence on $J$, while the spectroscopic quadrupole moments follow a parabolic trend. These behaviors are consistent with the predictions of the PRM. Experimental data are scarce. We find that the magnetic moment and spectroscopic quadrupole moment for the ground state $J^\pi = 5/2^+$ are $\mu = 0.32~\mu_N$ and $Q_s = 299~e\,\mathrm{fm}^2$, respectively, in good agreement with the corresponding data $\mu = 0.360(7)~\mu_N$ and $Q_{s} = 311(6)~e\,\mathrm{fm}^2$, respectively. For the isomeric state $3/2^+$, we predict $\mu = -0.48~\mu_N$ and $Q_s = 171~e\,\mathrm{fm}^2$, compared with the experimental values $\mu = -0.37(6)~\mu_N$ and $Q_{s} = 174(6)~e\,\mathrm{fm}^2$, respectively. Overall, the available data for the magnetic and spectroscopic quadrupole moments of several low-lying states in $^{229}$Th are reasonably well reproduced by the MR-CDFT. In addition, we have calculated the magnetic moments of the first and second $5/2^-$ states in $^{229}$Th, as shown in Fig.~\ref{fig:Th229_Pa229_Level}. The obtained values are $ \mu(5/2_1^-)=0.31~\mu_N$ and $ \mu(5/2_2^-)=-0.21~\mu_N$, respectively. The noticeable difference between these magnetic moments indicates that the two states originate from distinctly different quasiparticle configurations.

The spectroscopic quadrupole moments of the positive-parity bands with $K=5/2$ and $K=3/2$ in $^{229}$Pa are shown in Fig.~\ref{fig:Pa229_Qs}, where the predictions by the PRM are also given for comparison. One can see that with a proper adjustment of the parameters in the PRM, one is able to reproduce the results of MR-CDFT. The detailed values for the magnetic dipole moments $\mu$ and spectroscopic quadrupole moments $Q_s$ of $^{229}$Pa are provided in Table~\ref{tab:Th229}. One finds the ground state with $\mu(5/2^+)=2.26 \mu_N$ and  $Q_s(5/2^+)=319~\mathrm{efm}^2$. The parity doublet shares nearly identical values $\mu(5/2^-)=2.21,\mu_N$ and $Q_s(5/2^-)=318~\mathrm{efm}^2$.   A similar situation is found for the parity doublets with $J^{\pi}=3/2^{\pm}$ and $K=3/2$.

\begin{figure}[]
 \centering
\includegraphics[width=8.5cm]{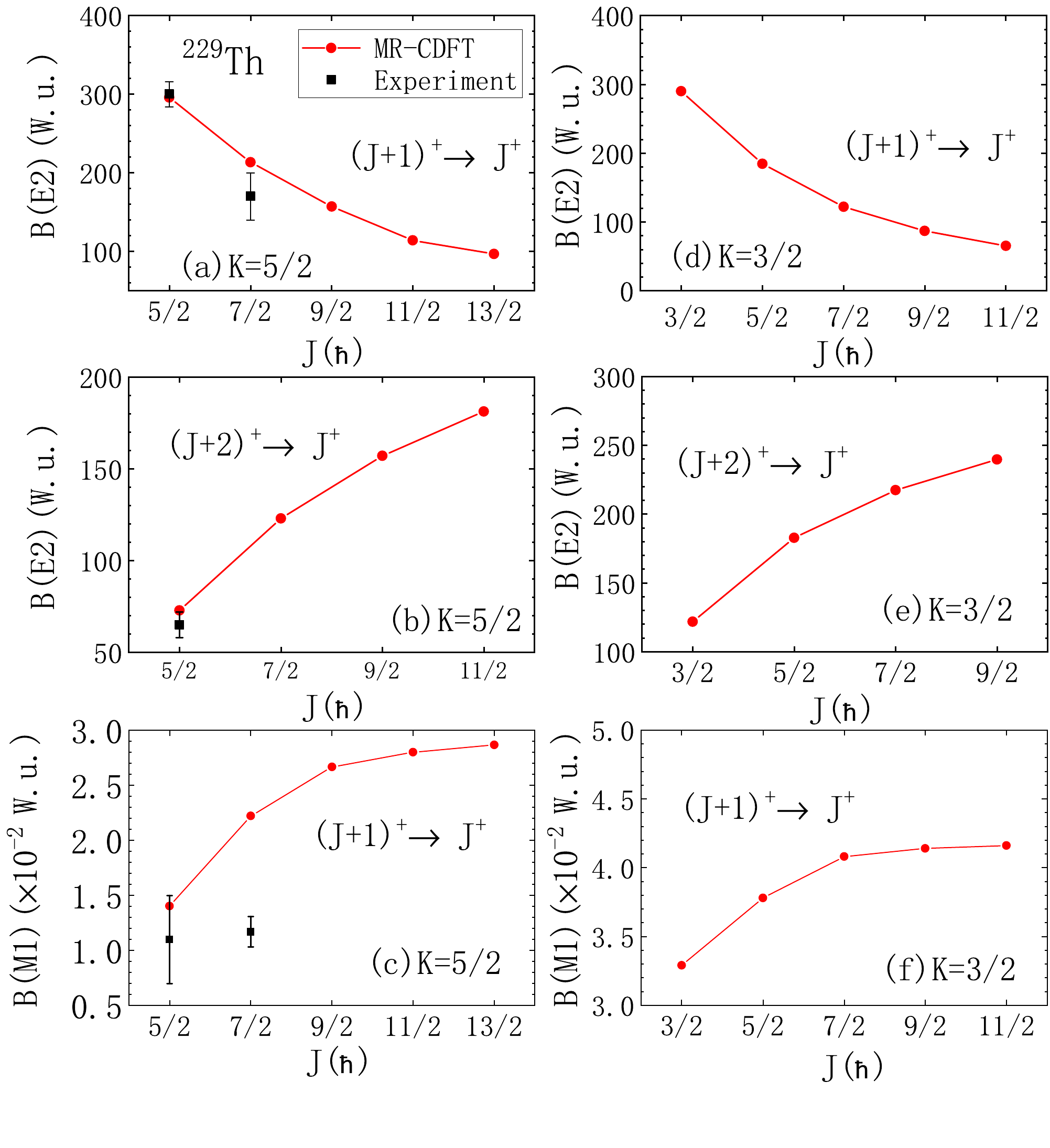}
\caption{ 
Same as Fig.~\ref{fig:Xe129_trans}~(a–c), but for $^{229}$Th. Panels (a–c) correspond to the positive-parity $K=5/2$ band, while panels (d–f) correspond to the positive-parity $K=3/2$ band. The experimental data are taken from Refs.~\cite{NNDC,Minkov:2017kju,Chen:2025rlz}.
}
 \label{fig:Th229_E2_M1}
 \end{figure}

Figure~\ref{fig:Th229_E2_M1}  displays the comparison of the calculated intraband  $E2$ and $M1$ transition strengths for the ground-state  and excited bands of $^{229}$Th,  with available experimental data. It is shown that the transition strengths of experimental data are reproduced rather well. The $B(M1; 9/2^+\rightarrow 7/2^+)$  is slightly overestimated in \nuclide[229]{Th}.  Overall, both bands display similar spin dependence, i.e., $B(E2)$ increases with $J$ for $\Delta J=2$ transitions, and  decreases for $\Delta J=1$, while $B(M1)$ exhibits a parabolic variation with $J$. All these behaviors are consistent with the predictions of PRM for $K\neq1/2$ bands.

    \begin{figure}[]
 \centering
\includegraphics[width=8.5cm]{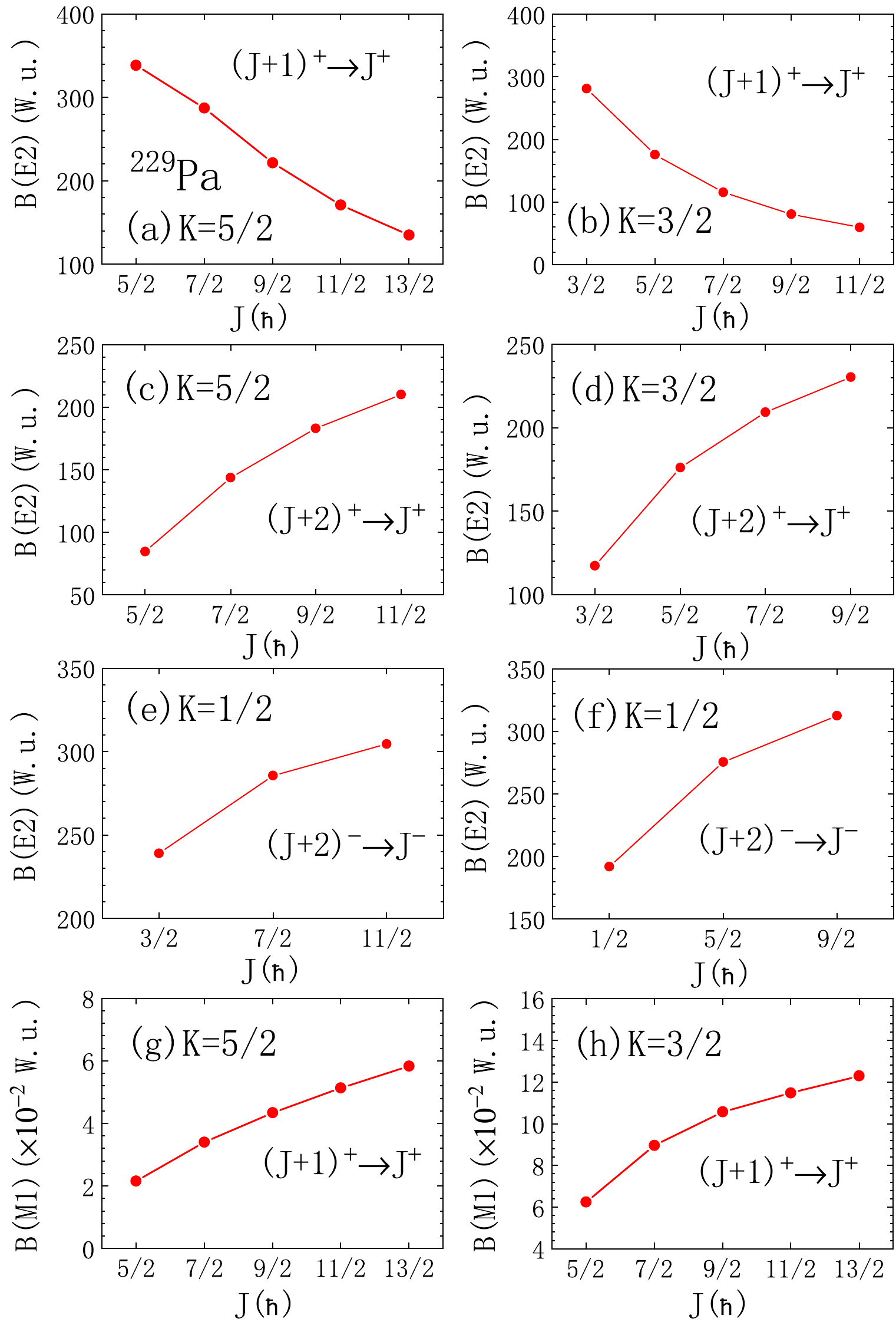}
\caption{ Same as Fig.~\ref{fig:Th229_E2_M1}, but for $^{229}$Pa. (a,c,g) with $K=5/2$; (b,d,h) $K=3/2$; and (e,f) $K=1/2$.
}
 \label{fig:Pa229_E2_M1}
 \end{figure}

Figure~\ref{fig:Pa229_E2_M1} display the  $B(E2)$ and $B(M1)$ transition strengths in \nuclide[229]{Pa} as a function of angular momentum  $J$. The absolute values of the $E2$ transition strengths in \nuclide[229]{Pa} are systematically larger than those in \nuclide[229]{Th} because the low-lying states of \nuclide[229]{Pa} with $K=5/2$ are dominated by the configurations with larger quadrupole deformation than those of \nuclide[229]{Th}, see Fig.\ref{fig:Th229_Pa229_PE}.  Again, the  behaviors of the  $B(E2)$ and $B(M1)$ transition strengths are consistent with the predictions of PRM based on a strongly  quadrupole-octupole deformed shape.

\begin{table}[bt]
\centering
\tabcolsep=20pt
\caption{The theoretical $B(E3)$ values (W.u.) of the intraband $E3$ transitions for the $K=5/2$ band of $^{229}$Th and  $^{229}$Pa from the MR-CDFT calculation.
}
\begin{tabular}{ccc}
  \hline \hline \\
    $J^{\pi_i}_i\to J^{\pi_f}_f$  & \nuclide[229]{Th} & \nuclide[229]{Pa} \\ 
  \hline
   $7/2^-_1\rightarrow 5/2_1^+$  & 191 & 183 \\ 
   $9/2^-_1\rightarrow 7/2_1^+$  & 33 & 30 \\ 
   $11/2^-_1\rightarrow 9/2_1^+$  & 0.4 & 0.3 \\ 
   $13/2^-_1\rightarrow 11/2_1^+$  & 5.4 & 6.5 \\ 
   $15/2^-_1\rightarrow 13/2_1^+$  & 15 & 21 \\ 
   $11/2^-_1\rightarrow 5/2_1^+$ & 42  & 39 \\ 
   $13/2^-_1\rightarrow 7/2_1^+$ & 75  & 71 \\ 
   $15/2^-_1\rightarrow 9/2_1^+$ & 95  & 96 \\
   \hline \hline
\end{tabular}
\label{tab:ThPa}
\end{table}

    \begin{figure}[]
 \centering
\includegraphics[width=8.5cm]{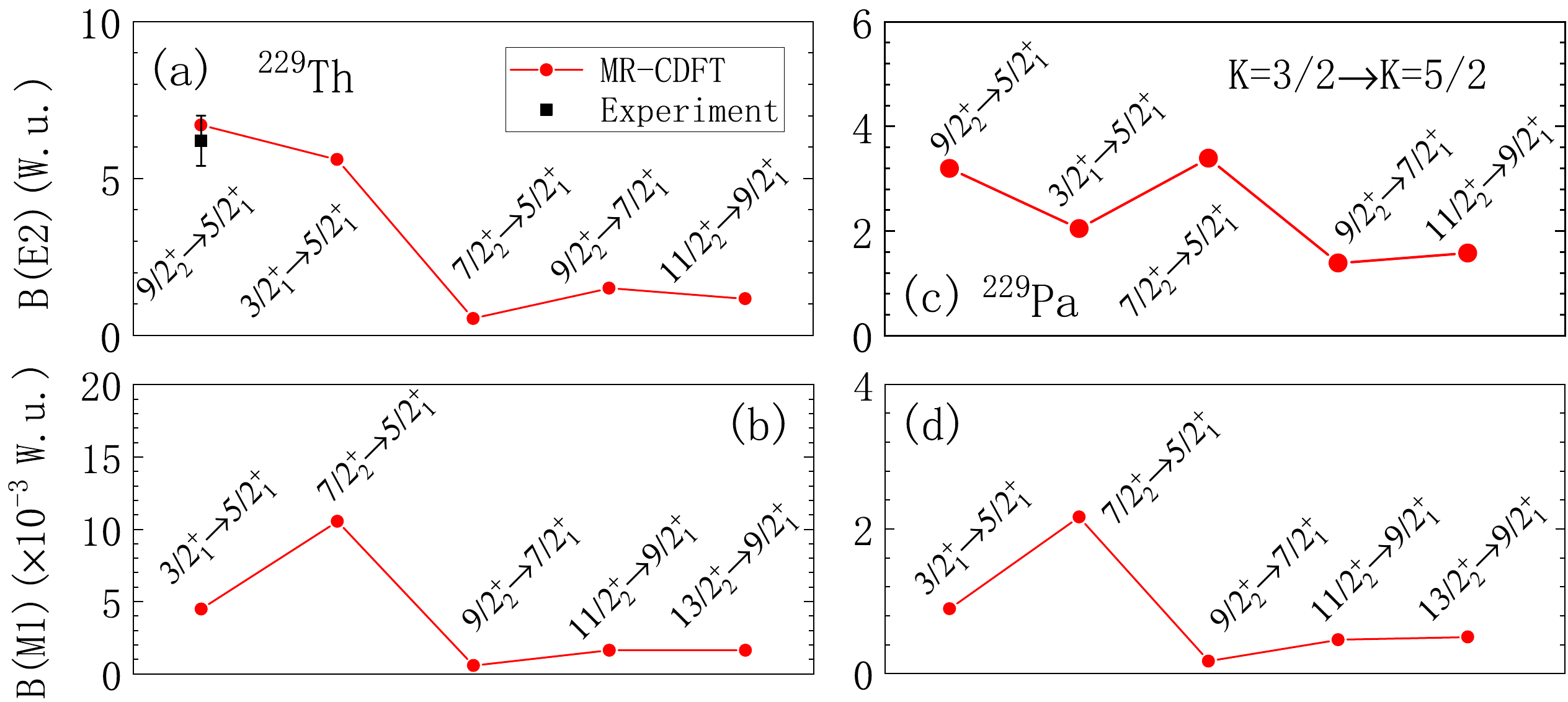}
\caption{
Electric quadrupole transition strengths $B(E2)$ and magnetic dipole transition strengths $B(M1)$ between the low-lying excited band $(K=3/2)$ and the ground-state band $(K=5/2)$ in $^{229}$Th (a), (b) and $^{229}$Pa (c), (d), calculated with the MR-CDFT.
 The data is taken from~\cite{NNDC,Minkov:2017kju,Chen:2025rlz}
}
 \label{fig:Th229_Pa229_E2_M12_13}
 \end{figure}

Table~\ref{tab:ThPa} summarizes the calculated $B(E3)$ values for the intraband transitions between the states of the $K = 5/2$ band in $^{229}$Th and $^{229}$Pa obtained from the MR-CDFT calculations. It is seen that the $B(E3)$ values of these two nuclei are comparable, again indicating the similarity of the predominant configurations in the ground-state bands of both nuclei. Quantitatively, the $B(E3; 7/2^-_1 \rightarrow 5/2^+_1)$ values are about 190~W.u. in both $^{229}$Th and $^{229}$Pa, roughly three times larger than that found in \nuclide[225]{Ra} (see Fig.~\ref{fig:Ra225_spectra}). 

The $B(E2)$ and $B(M1)$ values for the interband transitions between the states of the $K = 3/2$ and $K = 5/2$ bands in $^{229}$Th and $^{229}$Pa are shown in Fig.~\ref{fig:Th229_Pa229_E2_M12_13}. We note that the mixing of quasiparticle configurations with different $K$ quantum numbers plays a decisive role in describing the interband transitions. The evolution patterns of both $B(E2)$ and $B(M1)$ in $^{229}$Pa with increasing angular momentum $J$ are similar to those found in $^{229}$Th. Quantitatively, we obtain $B(E2; 9/2^-_2 \rightarrow 5/2^+_1) = 3.2$~W.u., $B(E2; 7/2^-_2 \rightarrow 5/2^+_1) = 2.1$~W.u., and $B(M1; 3/2^-_1 \rightarrow 5/2^+_1) = 0.00098$~W.u. in \nuclide[229]{Pa}, which are systematically weaker than the corresponding theoretical values in \nuclide[229]{Th}, namely 6.7~W.u., 5.6~W.u., and 0.0045~W.u., respectively. We further note that the interband transition $B(E2; 9/2^+_2 \rightarrow 5/2^+_1) = 6.7$~W.u. in \nuclide[229]{Th} is in excellent agreement with the experimental value of 6.2(8)~W.u. For the $B(M1; 3/2^+_1 \rightarrow 5/2^+_1)$ transition, our calculated value for \nuclide[229]{Th} is 0.0045~W.u., several times smaller than the experimental estimates of 0.017–0.03~W.u. deduced from the measured lifetime of the isomer~\cite{Chen:2025rlz}, assuming a purely $M1$ radiative decay. Moreover, the $B(E2)$ and $B(M1)$ transition strengths from higher excited states are found to be very small, indicating that interband transitions are strongly hindered in both $^{229}$Th and $^{229}$Pa.

\section{Summary}   
\label{sec:summary}

In this paper, we have employed the multireference covariant density functional theory (MR-CDFT) to systematically investigate the low-lying structure and spectroscopic properties of nuclei that are of particular interest in electric dipole moment (EDM) studies, including $^{129}$Xe, $^{199}$Hg, $^{225}$Ra, $^{229}$Th, and $^{229}$Pa. The MR-CDFT framework incorporates angular-momentum, parity, and particle-number projections on top of symmetry-broken mean-field states. It also includes generator coordinate method (GCM) calculations with simultaneous mixing of configurations with different quadrupole ($\beta_2$) and octupole ($\beta_3$) deformations. This fully microscopic approach does not rely on any phenomenological effective charges, allowing for parameter-free and quantitative predictions of spectroscopic observables.

For $^{129}$Xe and $^{199}$Hg, our MR-CDFT calculations reveal pronounced octupole-vibrational characteristics. The potential energy surfaces of both nuclei are soft with respect to octupole deformation, even when the quadrupole deformation remains small. When the octupole degree of freedom is included, the calculated $E2$ and $M1$ transition strengths, as well as the spectroscopic quadrupole and magnetic dipole moments of the low-lying states, are reasonably reproduced, even though the spectra are somewhat overall overestimated. These results further confirm that the inclusion of octupole correlations is essential for an accurate microscopic description of the low-energy spectra and electromagnetic properties of these nuclei.

For $^{225}$Ra, $^{229}$Th, and $^{229}$Pa, the MR-CDFT calculations provide a unified microscopic description of the nearly degenerate parity-doublet rotational bands, revealing the crucial role of octupole correlations in the nuclei of this mass region.
\begin{itemize}
\item For $^{225}$Ra, the mean-field energy minimum at $(\beta_2, \beta_3) = (0.18, 0.11)$ corresponds to a rather strong octupole-deformed nucleus. The MR-CDFT reproduces the energy of the first opposite-parity state, $E_x(1/2^-) = 78~\mathrm{keV}$, in good agreement with the experimental value of 55~keV. The ground-state rotational band exhibits a clear $\Delta J = 2$ pattern, with $B(E2)$ strengths for $\Delta J = 2$ transitions significantly larger than those for $\Delta J = 1$, confirming the robust rotational character and rather strong octupole correlations. The two factors, i.e., the  ground state has the spin–parity $1/2^+$ and thus zero electric quadrupole moment, and the existence of a low-lying $1/2^-$ state, make $^{225}$Ra particularly interesting for atomic EDM searches.  

\item  The mean-field energy minimum of $^{229}$Th is found at $(\beta_2, \beta_3) = (0.22, 0.15)$, indicating that this nucleus exhibits an even stronger octupole deformation than \nuclide[225]{Ra}. The MR-CDFT calculations successfully reproduce the known low-lying states, including the ground-state band ($J^\pi = 5/2^+$) and the isomeric state ($J^\pi = 3/2^+$) at $E_x \approx 23~\mathrm{keV}$. The predicted spectroscopic quadrupole and magnetic dipole moments, as well as the $B(E2)$ transition strengths, are in good agreement with data. In addition, we predict a nearly degenerate negative-parity band as the parity partner of the ground-state band, arising from strong octupole 
deformation, and a weakly deformed $K = 3/2^-$ band with limited rotational collectivity. The very low excitation energy of the bandhead state $J^\pi = 5/2^-$ at $E_x \approx 15~\mathrm{keV}$ makes \nuclide[229]{Th} a promising candidate for exhibiting a significantly enhanced Schiff moment~\cite{Zhou:2025_letter}.

\item For $^{229}$Pa, which differs from $^{229}$Th by the conversion of one neutron into a proton, the calculations predict a similar low-lying structure: a ground state with $J^\pi = 5/2^+$ and two nearly degenerate excited states, $J^\pi = 5/2^-$ and $3/2^+$, with excitation energies of 16~keV and 32~keV, respectively. Although experimental data are scarce, the predicted strong octupole correlations, large quadrupole moments, and rotational patterns analogous to those of $^{229}$Th make $^{229}$Pa another interesting candidate for future experimental studies, even though the ground-state spin-parity of both nuclei are $5/2^+$ with nonzero electric quadrupole moments. 
\end{itemize}

In summary, the MR-CDFT framework provides a comprehensive, parameter-free description of low-lying nuclear structures across the studied nuclei. The calculations underscore the essential roles of octupole correlations and shape mixing in reproducing spectroscopic observables, rotational bands, and electromagnetic transition strengths. For experimentally less explored nuclei such as $^{229}$Pa, the present results offer microscopic predictions of low-lying spectra and transition strengths, thereby providing valuable guidance for future measurements. It is worth noting that extending the current framework to incorporate triaxial deformation and noncollective excitation configurations may further refine the description of nuclei such as $^{129}$Xe and $^{199}$Hg. However, such extensions are not expected to lead to significant changes in the predictions for the well deformed $^{225}$Ra, $^{229}$Th, and $^{229}$Pa.

 \section*{Data availability statement}
The data that support the findings of this article are openly available \cite{Data}.

\section*{Acknowledgments} 
We thank J. Engel, H. Hergert, J. Meng, J. T. Singh, and Y. B. Wu for helpful discussions, and Q. L. Feng and C. F. Jiao for carefully reading the manuscript. This work was supported in part by the National Natural Science Foundation of China under Grant Nos. 12405143 and 12375119. 

 
%

  \end{document}